\documentclass[twocolumn]{aastex63}

\usepackage{amsmath}
\usepackage{float}

\received{May 13, 2026}
\revised{July 10, 2026}
\accepted{July 21, 2026}
\submitjournal{JAAVSO}

\shorttitle{Variable Star Polarimetry with PICSARR-2}
\shortauthors{Cotton et al.}

\begin{document}

\title{Variable Star Polarimetry with PICSARR-2}

\correspondingauthor{Daniel V. Cotton} 
\email{dc@mira.org}

\author[0000-0003-0340-7773]{Daniel V. Cotton}
\affiliation{Monterey Institute for Research in Astronomy, 200 Eighth Street, Marina, CA 93933, USA.}

\author[0000-0002-5726-7000]{Jeremy Bailey}
\affiliation{School of Physics, University of New South Wales, Sydney, NSW 2052, Australia.}
\affiliation{Western Sydney University, Locked Bag 1797, Penrith-South DC, NSW 2751, Australia.}

\author[0009-0000-7505-9072]{Logan Barrett}
\affiliation{Mountain View High School, 3535 Truman Avenue, Mountain View, CA 94040, USA.}
\affiliation{Monterey Institute for Research in Astronomy, 200 Eighth Street, Marina, CA 93933, USA.}

\author{Glenn Henderson}
\affiliation{Monterey Institute for Research in Astronomy, 200 Eighth Street, Marina, CA 93933, USA.}

\author{Kim Sumagang}
\affiliation{Monterey Institute for Research in Astronomy, 200 Eighth Street, Marina, CA 93933, USA.}

\author[0009-0004-7192-0055]{Eric C. Haase}
\affiliation{Monterey Institute for Research in Astronomy, 200 Eighth Street, Marina, CA 93933, USA.}


\begin{abstract}

We describe the upgraded Polarimeter using Imaging CMOS Sensor and Rotating Retarder 2 (PICSARR-2), describe its applications, and characterize its performance for stellar polarimetry on a 36-inch and 14-inch telescope.  On the larger telescope in the SDSS $g^\prime$, $r^\prime$ and $i^\prime$ filters a precision of $\sigma_p=$ 5.7 ppm on bright stars is recorded using a fast modulation rate corresponding to frame exposures of 12 ms; accounting for the internal errors in the individual observations gives a limiting precision of $e_p=$ 1.3 ppm. Longer frame rates are required for stars with $m > 5$, but the recorded errors only underperform a photon shot noise derived extrapolation for $m > 8$, when even longer frame exposures are required. Stars as faint as $m = 11$ were observed. The position angle precision is measured as 0.0845 degrees, and there is very good agreement between observations made by both the PICSARR-2 and HIPPI-2 polarimeters. On the smaller telescope the instrument's performance approaches similar levels, and there is good cross-platform stability in the observation of standard stars. PICSARR-2 is therefore an excellent instrument to explore stellar variability due to a range of phenomena; examples from our ongoing pulsating star campaign and other variable star programs are presented.

\end{abstract}


\section{Introduction} 
\label{sec:intro}

Primarily, this paper describes a new high precision polarimeter built for small telescopes. However, with the push to add a Polarimetry Special Interest Group (SIG) to the AAVSO (Ignace, Priv. Comm.) we also aim to introduce stellar polarimetry, its instrumentation, and applications to an AAVSO audience. 

\subsection{Linear Polarimetry}
\label{sec:linear_pol}

Conceptually, linear polarization describes the average orientation of photons in a beam, where the orientation of each photon may be thought of as the position angle, perpendicular to the direction of travel, of the oscillation in the wave that describes it. If the photons are randomly aligned then the beam will be essentially unpolarized. If all of the photons are aligned the same way, the beam is polarized, with $p=100\%$, at the preferred position angle, $\theta$. If there are a mix of photon alignments then the beam is partially polarized, with $p$ and $\theta$ given by the vector sum of the constituent photons.

\begin{figure*}[t!]
\includegraphics[width=\textwidth]{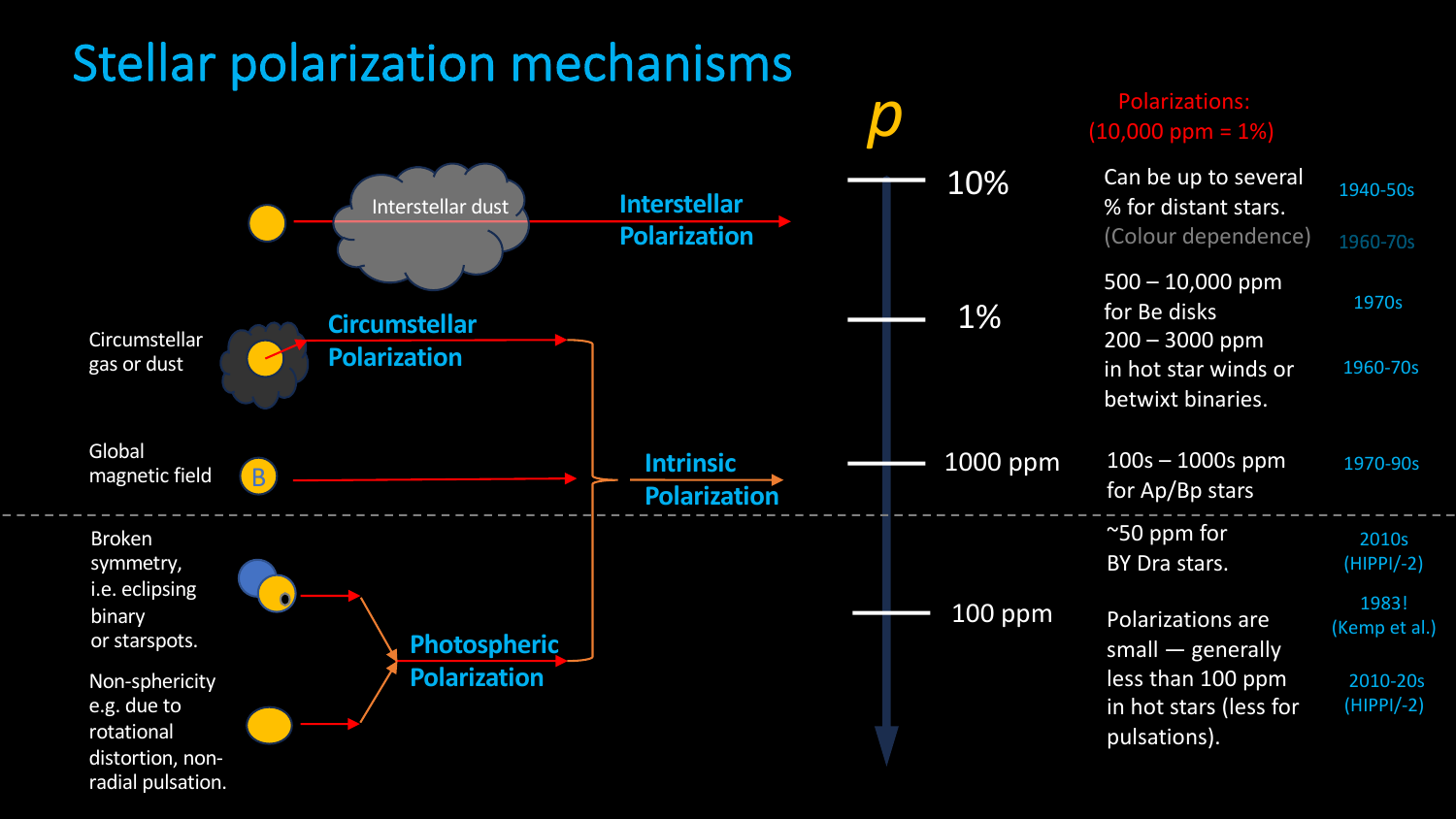}
\caption{A summary of selected polarization mechanisms, the era of their initial discovery and investigation, and the typical polarizations associated with them.}
\label{fig:mechanisms}
\end{figure*}

For convenience, astronomers often work with Stokes parameters, $Q$ and $U$, which are the vector components of polarized flux \citep{Stokes1851}. Unlike the $x$ and $y$ vector components of projectile velocity, for instance, the Stokes parameters are not perpendicular, but separated by 45 degrees. This is because polarization is only a pseudo vector, with orientation but no direction. Thus, in the Equatorial system, 0 degrees (North) corresponds to Stokes $Q$ and 45 degrees to Stokes $U$; 90 degrees (East) is $-Q$ and 135 degrees is $-U$. The total flux is usually labelled as Stokes $I$, and the normalized Stokes parameters given the lowercase characters, \begin{equation}q=\frac{Q}{I},\ \ u=\frac{U}{I}.\end{equation} Typically, one first gets the Stokes parameters from an instrument by making intensity measurements of the polarized components at four position angles, so that \begin{equation}q=\frac{I_{0}-I_{90}}{I_{0}+I_{90}}, \ \ u=\frac{I_{45}-I_{135}}{I_{45}+I_{135}},\end{equation} where the subscripts indicate the position angles, then converts to $p$ and $\theta$. To convert between the two systems: \begin{equation}p=\sqrt{q^2+u^2},\ \ \theta=\frac{1}{2}\arctan{\left(\frac{u}{q}\right)}.\end{equation} In practice, it is necessary to use the \textsc{arctan2} function -- commonly found in spreadsheet programs and computer languages -- in place of $\arctan$ in order to avoid an error of sign in some cases.

\subsection{Stellar Polarimetry}
\label{sec:stellar_pol}

Polarimetry is a valuable tool in variable star studies, but one that has been under-utilized. Stars are net linearly polarized when their light is scattered asymmetrically, either from the stellar atmosphere itself -- \textit{photospheric polarization} -- or from the gas and dust that surrounds them -- \textit{circumstellar polarization}. A net linear polarization may also arise due to magnetic fields or, in high energy environments, synchrotron processes. Polarimetry is particularly good at revealing details of geometry in unresolved systems, as well as particle size and other properties, where conventional techniques are inadequate. A comprehensive review of polarigenic mechanisms discovered to \citeyear{Clarke2010} is given by \citeauthor{Clarke2010}; in this section, and summarised in Fig. \ref{fig:mechanisms}, we highlight a few most applicable to broadband linear polarimetry of stars, to give a sense of the field.

\begin{figure}
\includegraphics[trim={-1cm, 3cm, 1cm, 1cm}, clip, width=0.975\columnwidth]{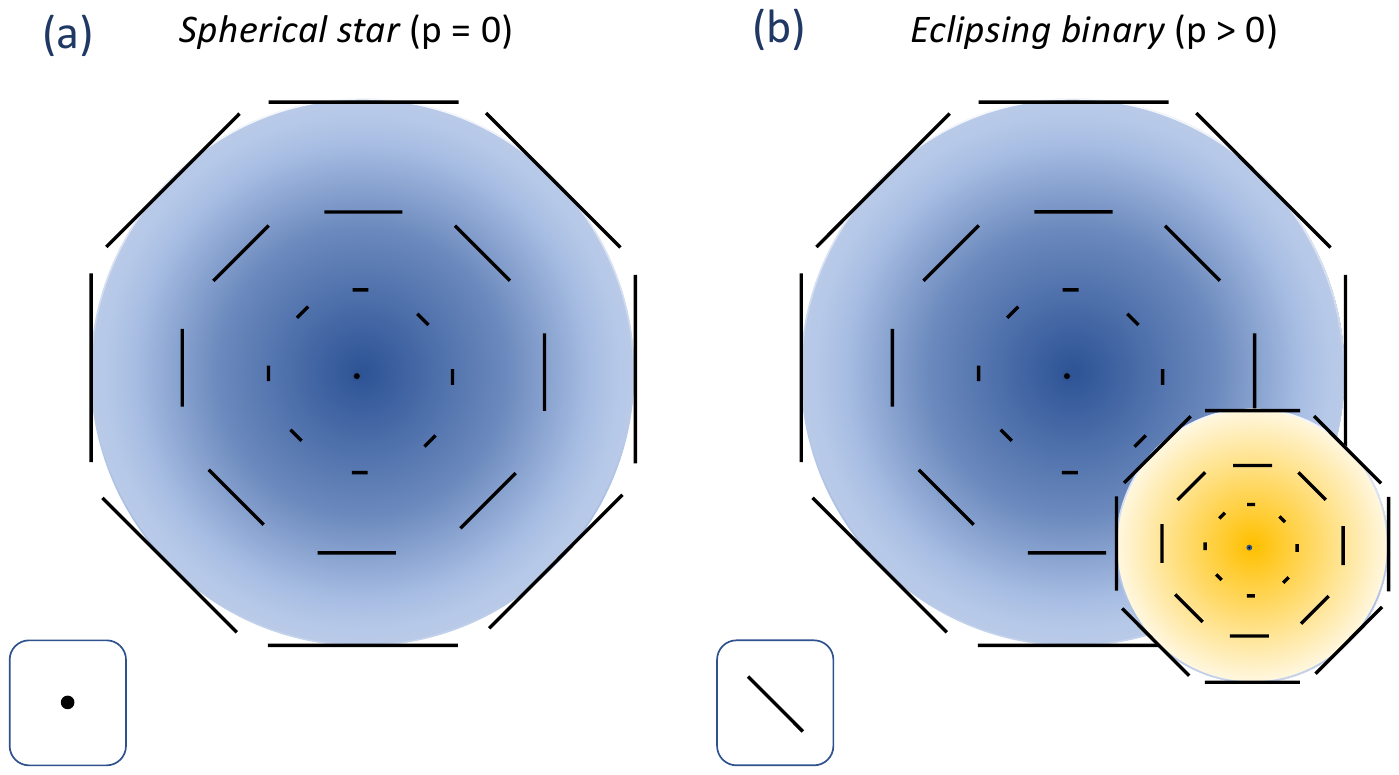}
\caption{Two star systems with different polarizations from an observer's perspective. The black lines are polarization vectors, in (a) where a spherical star is represented, the vector sum results in zero polarization, in (b) there is a net polarization resulting from a binary companion blocking part of the light from an eclipsed star -- this is the Chandrasekhar Effect. The square boxes at the bottom in this and subsequent figures show the net polarization from the observer's perspective.}
\label{fig:chandrasekhar}
\end{figure}

As an astronomical technique, polarimetry can trace its beginnings back more than 200 years (\citealp{Arago1811, Arago1855a, Arago1855b}). Yet, stellar polarimetry effectively began with the prediction by \citet{Chandrasekhar1946} of measurable photospheric polarization in eclipsing hot binary systems. The polarization of light emanating from a star is a function of its scattering angle, such that it increases toward the limb of the stellar disc. For a homogeneous spherical star, however, all the polarization vectors cancel out, resulting in unpolarized light (Fig. \ref{fig:chandrasekhar}a). In the scenario described by Chandrasekhar the spherical symmetry is broken by the eclipse (Fig. \ref{fig:chandrasekhar}b). The first polarimeters with photoelectric detectors were built to search for polarization produced this way. It was a long time before this mechanism was seen, but as instruments improved, other phenomena were discovered (Fig. \ref{fig:mechanisms}).

\subsubsection{Interstellar Polarization}
\label{sec:interstellar}

\begin{figure}
\includegraphics[trim={2cm, 3cm, 2cm, 1cm}, clip, width=\columnwidth]{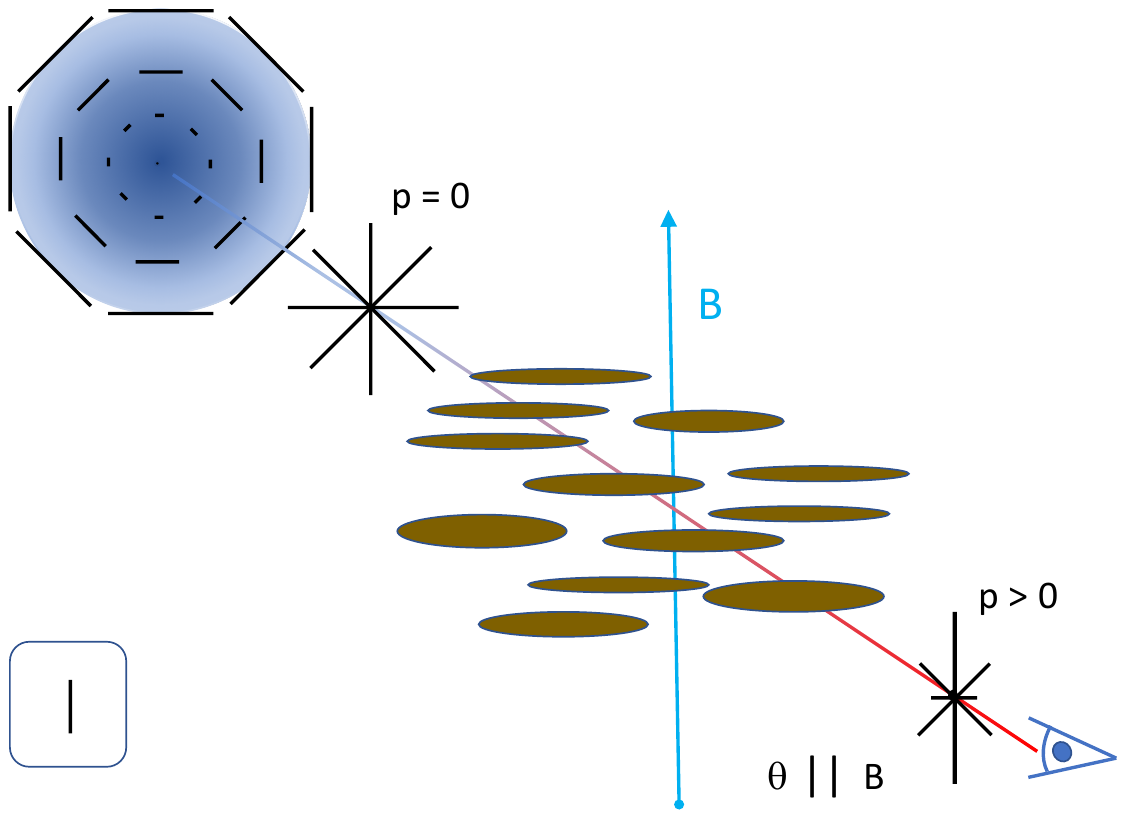}
\caption{A false-perspective depiction of the process of polarization by the interstellar medium (the eye represents the observer's view). Dust grains align perpendicular to the magnetic field, these then preferentially absorb light aligned with their long axis, resulting in a net polarization parallel to the magnetic (B) field. Interstellar polarization tends to be correlated with interstellar reddening because dust preferentially scatters blue light.}
\label{fig:interstellar}
\end{figure}

The first measurements of polarized starlight turned out to have nothing to do with the stars at all. Instead, polarization increasing with the distance to stars led to the discovery of \textit{interstellar polarization} \citep{Hiltner1949, Hall1949}. Aligned dust grains in the interstellar medium acted as a dichroic absorber preferentially absorbing photons of one orientation over those perpendicular \citep{Leverett1951}; this is shown in Fig. \ref{fig:interstellar}. In the 1950s and 60s astronomers used this discovery to map the Galactic magnetic field which acts to align the dust grains (e.g. \citealp{vanPSmith1956, Hall1958}) -- the long axis of oblate grains align perpendicular to the field, producing optical polarization parallel with the field. Later the wavelength dependence of interstellar polarization was found by \citet{Serkowski1975}. It was shown that this was related to the size distribution of the dust grains, and this used to learn about the evolution of interstellar dust within nebulae and in general (e.g. \citealp{Coyne1979, Whittet1996}).

For stellar astronomers, interstellar polarization can often be a confounding factor in their study of intrinsic polarization -- examples of which follow. The strategy to use to disentangle the interstellar component depends on the intrinsic mechanism. The interested reader is again referred to \citet{Clarke2010}, or for a briefer account, \citet{Ignace2025}.

\subsubsection{Circumstellar Polarization}
\label{sec:circumstellar}

As instrumental sensitivity improved it became possible to study the gas within individual star systems. Much of the 1960s and 70s was devoted to understanding the structures of circumstellar polarization, which in early-type stars is largely due to electron scattering from asymmetrically distributed hydrogen gas. When a photon is scattered it is polarized perpendicular to the scattering plane, i.e. the plane formed by the source (star), scattering center (circumstellar gas) and the observer. Where the gas is thin, \textit{single scattering} results -- i.e. most photons seen by the observer are scattered only once -- and there is no wavelength dependence; when the gas is thicker, there is wavelength dependent \textit{multiple scattering} \citep{Halonen2013}.

\begin{figure}
\includegraphics[trim={1cm, 5cm, 1cm, 1cm}, clip, width=\columnwidth]{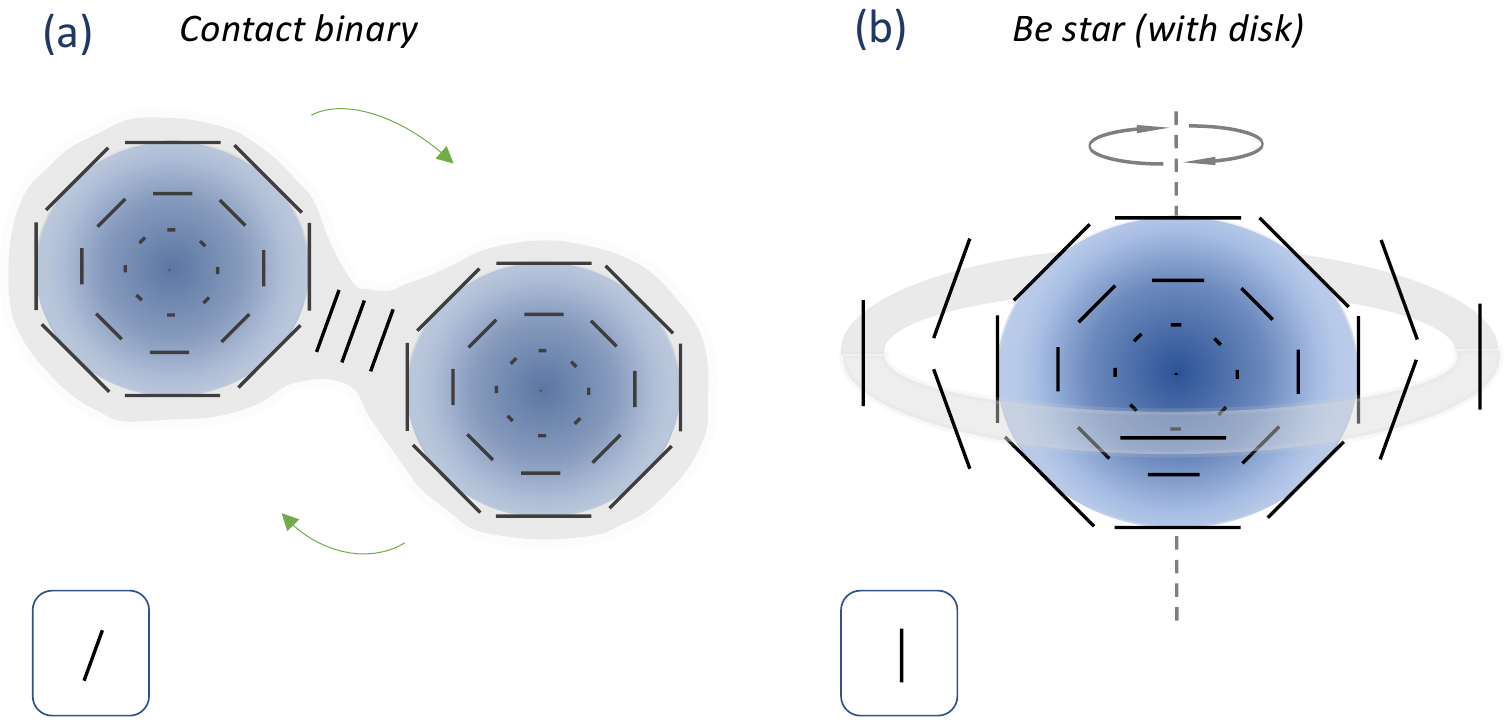}
\caption{Two examples of circumstellar polarization, where the distribution of hydrogen gas in the system is non-symmetric and produces polarization revealing geometry. (a) In a contact binary scattering from the gas entrained between the stars results in polarization revealing orbital parameters, (b) a Be star with a disk tilted below the plane of the observer is shown; the net polarization reveals the position angle of the disk on the plane of the sky. The density and geometry of the disk impacts the magnitude of $p$.}
\label{fig:circumstellar}
\end{figure}

The clumpy winds of early-type supergiants act as a scattering medium that produces large polarization at random or pseudo-random orientations \citep{Hayes1978, Hayes1986}. More recently a greater degree of structure has been revealed in the winds of Wolf-Rayet stars by polarimetry (e.g. \citealp{Ignace2023}). Phase-locked polarization, almost universally with two polarization peaks per orbit, is produced in a variety of close binary systems \citep{Manset2005}. A simple example is the gas entrained between the components of a compact binary acting as a scattering medium, as in Fig. \ref{fig:circumstellar}(a). In many cases the model of \citet{Brown1978} can be used to determine the orbital inclination, its position angle on the sky; and give details of the gas density and asymmetry (e.g. \citealp{Berdyugin2018, AbdulQadir2023}).

The emission regions around Be stars are confined to equatorial decretion disks; in these systems the polarization is perpendicular to the plane of the disk \citep{Yudin2001, Wisniewski2010}; an example is shown in Fig. \ref{fig:circumstellar}(b). The polarization magnitude is proportional to the quantity of gas. The formation and evolution of these disks is an active area of study for polarimetry \citep{CastanonEsteban2024, Carciofi2025}.

Variable circumstellar polarization can also arise from dusty environments; for example in the extended atmospheres of red giants \citep{McCall1980} and \mbox{T Tauri} stars \citep{Hough1981, Manset2002}. The mechanisms at work in these instances can be complex: whilst dust strongly polarizes, molecular emission or absorption results in depolarization of the continuum in affected regions \citep{Dinh-V-Trung2022}. The asymmetry that produces a net polarization might be caused by the morphology or heterogeneity of the dust or molecular cloud itself, asymmetric illumination due to stellar hot spots, pulsations, or asymmetrical obscuration of the stellar photosphere by the dust. 

Lastly, it is worth noting that small constant polarizations arise in the integrated light of dusty debris disk systems \citep{Marshall2023}. Where an accurate interstellar subtraction can be carried out, it is possible to discern the disk geometry: for a symmetric disk the polarization position angle will always be aligned to one of the two axes of the ellipse projected by the debris disk on the plane of the sky. Whether it is the major or minor axis depends on the dust grain properties and wavelength band. Multi-band polarimetry can be combined with data from other techniques to determine the debris dust properties \citep{Marshall2020}.

\subsubsection{Polarization by Magnetic Fields}
\label{sec:magnetic}

\begin{figure}[t!]
\includegraphics[trim={1cm, 1.5cm, 0.5cm, 1cm}, clip, width=\columnwidth]{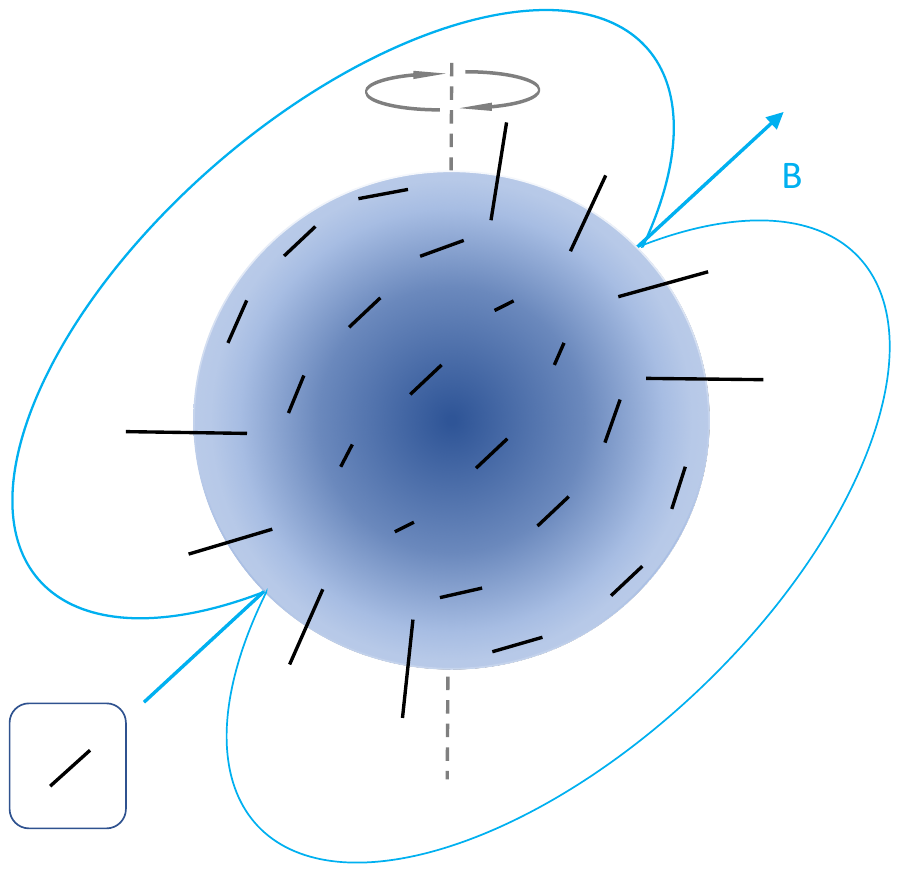}
\caption{Polarization from a magnetic star. Depicted here is a star with a dipole magnetic field presently aligned in the plane of the sky (and page), inclined by about 50 degrees from the rotational axis of the star. As the star rotates the poles will transit across the face, and the net field will change in both intensity and position angle. }
\label{fig:rotation}
\end{figure}

Another type of intrinsic polarization explored with conventional polarimeters is that due to stellar magnetism. Fairly large polarizations can be produced in Ap and Bp stars by the combination of line blanketing and the anomalous Zeeman effect, leading to an asymmetry correlated with the magnetic field geometry of the star (e.g. \citealp{Leroy1996}). The polarization varies as the star rotates. Linear stellar polarimeters were used for the preliminary work in field mapping before the advent of spectropolarimetry (particularly using circular polarization) superseded it \citep{Wade2000}. The component of the magnetic field in the plane of the sky is called the transverse component; it manifests as linear polarization tangential to the field lines, resulting in a position angle for the net field parallel with the magnetic field.  By contrast, the longitudinal component is that along the line of sight, the direction of which results in left- or right-hand circular polarization. Lately, the polarization due to the smaller fields of late-type magnetic stars like BY Dra variables \citep{Cotton2017b, Cotton2019a} has become important for the confounding effect it has in exoplanet studies \citep{Bott2016, Bailey2021}.

\pagebreak
\subsubsection{Photospheric Polarization}
\label{sec:photospheric}

Photospheric polarization arises from breaking the symmetry of the observed photosphere of the star. In hot stars where electron scattering dominates, it is much smaller in magnitude than other mechanisms in the visible range, but can become large in the ultraviolet (e.g. \citealp{Collins1991, Jones2022, Harrington2025}). The eclipse effect described by \citet{Chandrasekhar1946} and depicted in Fig. \ref{fig:chandrasekhar} is one such mechanism; it has been observed only once \citep{Kemp1983}. A similar mechanism relates to starspots or hot spots breaking the flux-symmetry of the disk; the detail of the phase curve will be different in this case \citep{Yakobchuk2018}, and the associated magnetic field will dwarf the effect. 

In close binary systems, light may be reflected from the photospheres of each component -- similar to entrained gas, this mechanism also traces the orbit \citep{Bailey2019} and produces larger polarizations than other hot star photospheric mechanisms. The wavelength dependence in binary reflection depends on the spectral types of the components \citep{Cotton2020}, and so could be used to determine the type of the companion in a single line binary system.

\begin{figure}
\includegraphics[trim={1cm, 4.5cm, 0.5cm, 3cm}, clip, width=\columnwidth]{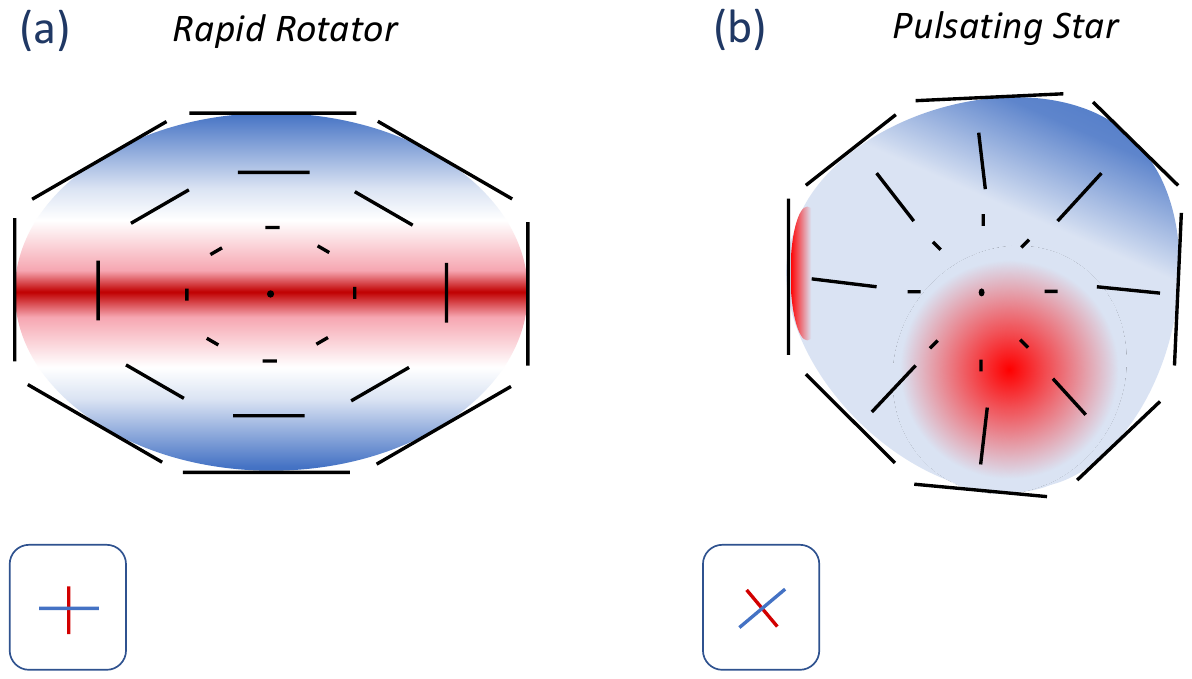}
\caption{Polarization in distorted stars: (a) Polarization from a rapidly rotating star. Here the effects of gravity darkening are greatly exaggerated to show how the polarization position angle depends on the wavelength of light. The position angle will flip from perpendicular to parallel to the rotation axis at longer wavelengths. The wavelength of the flip depends on the spectral type of the star. Note that the polarization magnitude is greatest at extreme UV wavelengths and \textit{generally} reduces with increasing wavelength. (b) A snapshot of a star pulsating in a non-radial mode. The complex shape, temperature pattern of the distorted star results in a wavelength dependent polarization that varies with the pulsation phase (adapted from a model created by Townsend, Priv. Comm.).}
\label{fig:rotation}
\end{figure}

Another way to produce a net photospheric polarization is through a distorted stellar disk. The simplest example being through rapid rotation resulting in an oblate spheroid \citep{Harrington1968, Cotton2017} as depicted in Fig. \ref{fig:rotation}. The net polarization magnitude is a combination of geometry (that is, where the asymmetry results from the rotation rate and the inclination of the rotation axis), wavelength, and something called the \textit{source function}, which describes the ratio of absorption to scattering in the photosphere. The rapid rotation mechanism produces constant polarization where the position angle is aligned perpendicular to the star's rotation axis at short wavelengths, and parallel to it at longer wavelengths. Larger polarizations are seen in hotter stars \citep{Cotton2017, Bailey2020a, Howarth2023} and those with higher luminosity \citep{Lewis2022, Bailey2024a}, so long as the photosphere is not obscured. 

Variable polarization is produced by photospheric distortion in pulsating stars \citep{Odell1979, Cotton2022a}. The polarigenic mechanisms are essentially the same as in rapid rotation, it's just that the shape of the star, the distribution of flux, etcetera is changing according to the pulsation mode. Radial and dipole modes produce no polarization owing to their symmetry, but in conjunction with photometry, it is possible to determine the mode degree, $\ell$, and azimuthal order, $m$, of other non-radial modes \citep{Watson1983, Cotton2022a}. Together, these parameters describe the distribution of vibration nodes on the surface of the star; determining them is crucial for asteroseismology. The polarization variation with phase traces out a sinusoid in $q$ and/or $u$. The polarization magnitude is partly dependent on the mode and inclination of the star, but will be larger for hotter and more luminous stars, and those with intrinsically larger amplitude pulsations \citep{Stamford1980}.

\subsection{Stellar Polarimeters}
\label{sec:polarimeters}

As in all technologically enabled endeavours, it is the continual improvement of instrumentation that has allowed progressively smaller polarimetric effects to be discovered. Depending on their intended application, polarimeters can be simple or complex -- employing many additional optical elements. For the sake of providing a background relevant to this work, we restrict this introduction to simple broadband instruments and what we consider the core innovations that progressed the art. This has the benefit of being more relevant to the development of polarimeters for small telescopes, for which there is interest for education \citep{Topasna2013, Kamara2024}, research \citep{Sinyavskii2013, Wolfe2015, Bailey2017, Bailey2023, Neilson2023}, or both \citep{Blay2022, Wiersema2023, Chai2026}.

\begin{center}
\begin{figure*}[t!]
\includegraphics[width=\textwidth]{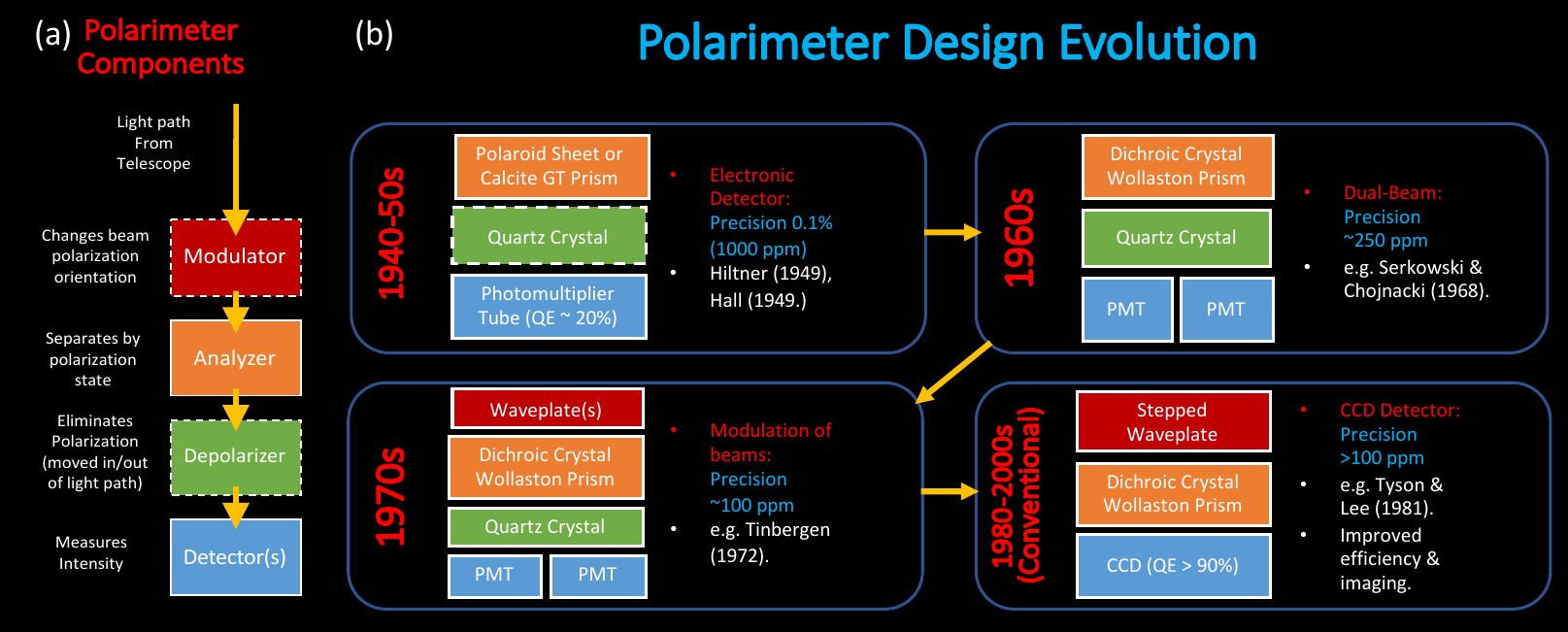}
\caption{(a) A block diagram showing polarimeter components and their typical arrangement. (b) The evolution of conventional polarimeter designs from the 1940s to the 2000s.}
\label{fig:pol_designs}
\end{figure*}
\end{center}

A good overview of the components of a polarimeter is given by \citet{Serkowski1974}. A schematic diagram of polarimeter components is shown in Fig. \ref{fig:pol_designs}(a). Every polarimeter consists of, at minimum, two basic components: a \textit{detector} which measures signal intensity and an \textit{analyzer} that separates one polarization orientation from the others. The analyzer always uses a dichroic or birefringent material\footnote{These are anisotropic materials where the orthogonal polarization vectors see different physical properties, e.g. absorbance (dichroism) or refractive index (birefringence), according to their alignment to the material's \textit{optic axis}. The optic axis is defined as the direction along which light can travel isotropically, i.e. it follows Snell's Law. If a ray of light enters a uniaxial crystal perpendicular to the optic axis, the component with its polarization vector perpendicular to the optic axis is called the ordinary ray (O-ray) and experiences refractive index $n_o$. Correspondingly, the component polarized parallel to the optic axis is the extraordinary ray (E-ray) and experiences refractive index $n_e$. The axis corresponding to the smaller refractive index is referred to as the fast-axis; whether this is $n_o$ or $n_e$ depends on the crystal properties.}.

To determine polarization in one Stokes parameter requires measurement of two orthogonal orientations (e.g. $I_{0}$ and $I_{90}$). If these are not measured simultaneously, then the signal must be modulated, either by rotating the analyzer, or using a dedicated \textit{modulator} to change the orientation of the signal. A further rotation/modulation by 45 degrees is required to measure the other Stokes parameter. Some instruments also employ a \textit{depolarizer} to negate any polarization sensitivity of the detector itself.

\begin{figure}[t!]
\includegraphics[trim={3cm, 4cm, 3cm, 2cm}, clip, width=\columnwidth]{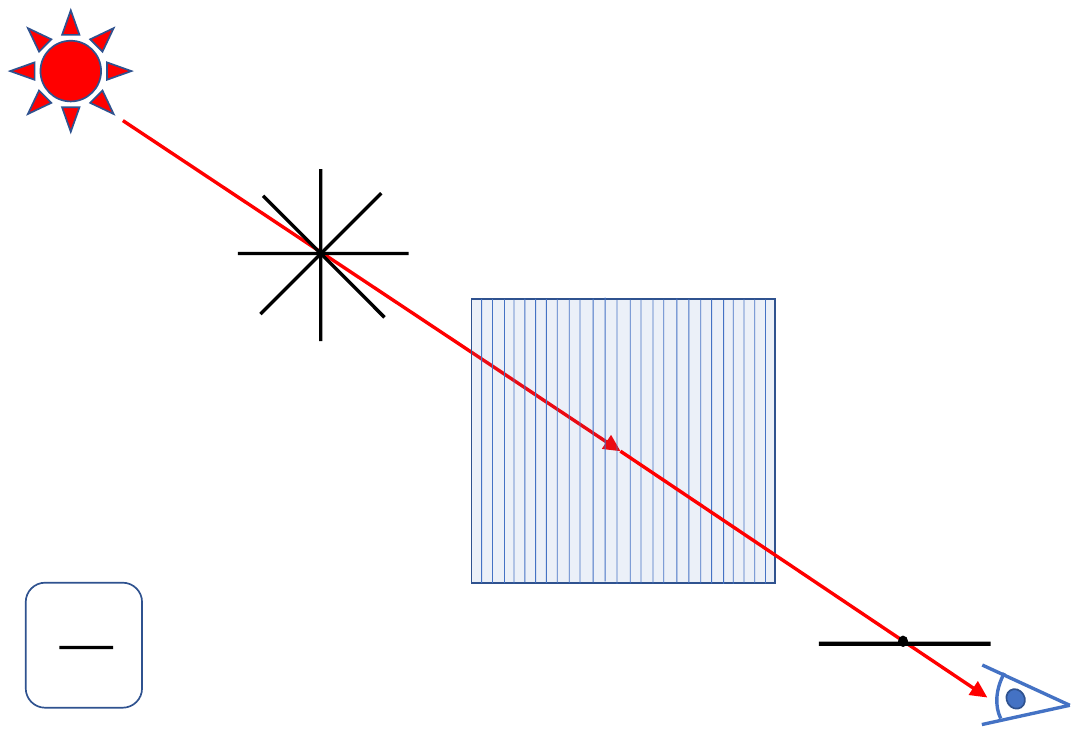}
\caption{False perspective view of a sheet of Polaroid -- a very effective polarization filter. Polaroid is akin to a wire-grid polarizer, where long chain polymers are drawn in one direction (vertically here). Light polarized parallel to the grid is absorbed, whereas light polarized perpendicular to it passes through.}
\label{fig:polaroid}
\end{figure}

\begin{figure}[t!]
\includegraphics[trim={3cm, 5cm, 7cm, 2cm}, clip, width=\columnwidth]{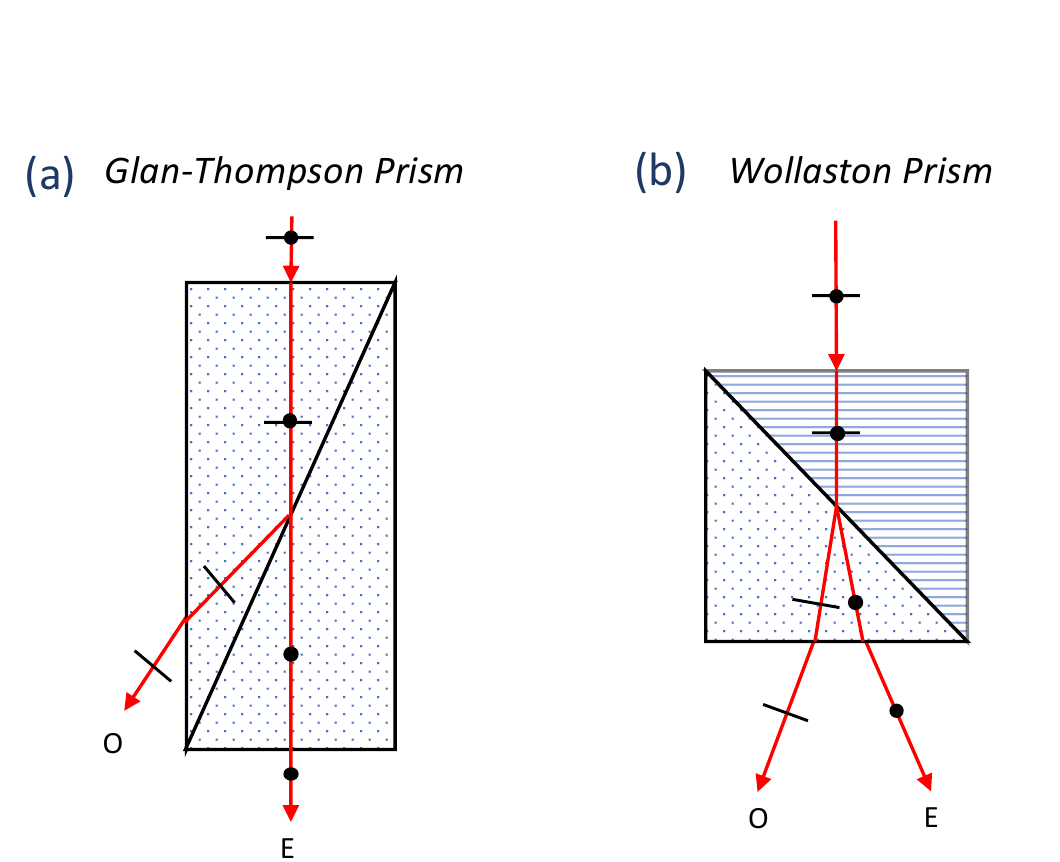}
\caption{Top down views of two polarizing beam splitters utilizing birefringent crystals (i.e. where the refractive index, $n$, depends on the direction of light travel through the crystal): (a) a Glan-Thompson prism, and (b) a Wollaston prism. The fill pattern depicts the optic axis direction -- the birefringence acts perpendicular to this axis, according to $n_o$ and $n_e$, causing the polarization components to travel at different speeds through the crystal and be reflected/refracted at different angles. Whereas in the Glan-Thompson prism two crystal wedges are arranged with aligned optic axes, the Wollaston prism opposes the two prisms. Both join the prisms with an optic cement. In the former it is the internally reflected beam that propagates at constant speed, and is therefore the ordinary (O) ray; the other beam is the extraordinary (E) ray. The O and E designations are more-or-less arbitrary for the Wollaston prism, as they swap at the join, here we choose relative to the exit prism.}
\label{fig:pbs}
\end{figure}

\subsubsection{Conventional Polarimeters}

The first electronic detector used for stellar polarimetry was the photo-multiplier tube (PMT), the 1P21 variety of which which has a best sensitivity (quantum efficiency, QE) of $\sim$20 percent. \citet{Hiltner1949} employed a polaroid sheet as an analyzer, while \citet{Hall1948, Hall1949} used a calcite Glan-Thompson prism analyzer and quartz crystal depolarizer. As shown in Fig. \ref{fig:polaroid}, a Polaroid sheet acts as a filter, passing a single polarization orientation. A quartz crystal is a birefringent material that has two axes with different refractive indices; the Glan-Thompson prism sandwiches two pieces together in an arrangement -- shown in Fig. \ref{fig:pbs}(a) -- that splits the beam into two components: a beam at high polarizing efficiency is transmitted in the forward direction, and a secondary beam is internally reflected at an acute angle to the direction of travel \citep{Glan1880, Thompson1881, Glazebrook1883, Thompson1886}. In \citeauthor{Hall1948}'s instrument the secondary beam was not measured (the O-ray is not as pure as the E-ray). Although also made of a birefringent material -- quartz in this case -- the depolarizer works by using two or more layers of different thicknesses arranged to scramble the polarization (c.f. \citealp{Lyot1929}). Since both instruments passed only one beam orientation, they relied on rotating the analyzer to modulate the signal; the precision of these set-ups was approximately 0.1 percent (1000 ppm).

The next major innovation was the \textit{dual-beam \mbox{polarimeter}}. By using a calcite Wollaston prism (\citealp{Wollaston1802, Wollaston1807}, Fig. \ref{fig:pbs}b) analyzer the two orthogonal components of polarization are passed at a small angular separation; they are then measured by two independent PMT detectors. This set-up halves the required observing time, and also improves precision. By simultaneously measuring the orthogonal polarization states, the effects of variable sky conditions and seeing noise, which acts to distort the beam at the detector, are largely negated \citep{Behr1961, Elvius1967} -- the exact improvement depends on how well matched the spatial sensitivity profiles of the two detectors are as a ratio of their overall sensitivity. The instrument still had to be rotated to measure the other Stokes parameter. \citet{Serkowski1969} were able to achieve a precision of 250 ppm with this arrangement. 

\begin{figure}[t!]
\includegraphics[trim={4.5cm, 4cm, 4.5cm, 3cm}, clip, width=\columnwidth]{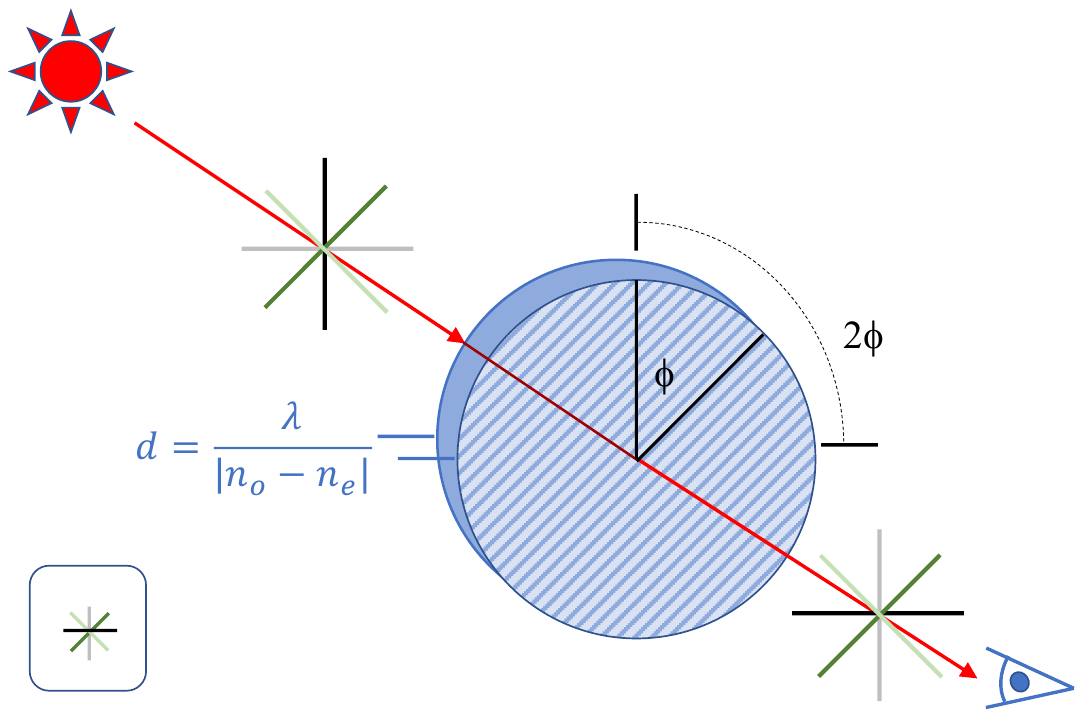}
\caption{False perspective view of a half-waveplate modulator. Shown is a zero-order, or single crystal, device, where the thickness is chosen such that the path length difference, $d$ for two orthogonal polarization vectors aligned parallel and perpendicular to the fast axis (here aligned to the optic axis) is half the wavelength, $\lambda$. This device rotates the plane of polarization by $2\phi$, where $\phi$ is the angle between the optic axis and the plane of polarization. More 
 complex, double-crystal and multi-crystal devices can be used to make the device achromatic or super-achromatic, respectively (c.f. \citealp{Pancharatnam1955}).}
\label{fig:hwp}
\end{figure}

A waveplate induces a phase shift between two orthogonal polarizations of light through birefringence, i.e. through a refractive index difference between the fast-axis and slow-axis of the device \citep{Fresnel1822}. One of the simplest modulators is a spinning half-waveplate -- where the phase shift is half a wavelength. Such a device rotates (modulates) the plane of polarization by $\theta=2\phi$, where $\phi$ is the angle between the polarization position angle of the incoming beam and the fast-axis of the crystal (Fig. \ref{fig:hwp}). The primary advantage of a modulator is that it cancels instrumental effects resulting from differences in the light path of the two orthogonal beams. In particular, if a different detector is used for each beam, then differences in their response curves can be cancelled. Using such a device \citet{Tinbergen1972} constructed a polarimeter with a precision a bit better than 100 ppm.

CCDs, which have peak QE's of $\sim$95$+$ percent, replaced PMTs as the detector of choice for most astronomy in the 1980s \citep{howell2006}. In the instrument of \citet{Tyson1981}, for instance, both beams were focussed onto different areas of the same detector. In this type of polarimeter, the area around the star images can be used as a simultaneous sky background, which helps with accuracy. Imaging data on resolved astrophysical objects or extended fields may also be taken. However, despite great improvements in efficiency, without schemes to distribute different components of polarization around the detector, the best precision of such instruments was no better than the previous generation, because seeing noise was the limiting factor \citep{Clarke2002}. CCDs are not capable of the fast read-out needed without generating excessive read noise, and so are a poor choice for bright star polarimetry in particular.

\subsubsection{High-Precision Polarimeters}
\label{sec:highprecpol}

According to \citet{Tinbergen1973}, to beat seeing noise limitations, a modulation rate of at least 10 Hz, and preferably 25 Hz or faster is necessary. This is generally a requirement of building a high-precision polarimeter -- one with a better precision than 100 ppm. Until recently, a polarimeter could either be efficient (i.e. in the QE sense) by using a CCD detector, or achieve high precision through other less efficient detector technologies. Consequently, high precision polarimeters tended to be exotic, and have some unique drawback that limited their use.

Consider, for instance, the most successful of these instruments, HIPPI \citep{Bailey2015} and HIPPI-2 \citep{Bailey2020}, which were used to discover three of the electron scattering photospheric polarization mechanisms \citep{Cotton2017, Bailey2019, Cotton2022a}. It achieves precisions of ~3-4 ppm -- compared to the 8 ppm of the next best instruments -- through the use of ferro-electric liquid crystal modulators and modern (43 percent peak-QE) photo-multiplier tube detectors. On a 1-m class telescope its operation is limited to stars with $m<10$ \citep{Cotton2022b}. The low detector efficiency along with significant observing overheads limits high-cadence studies to the brightest stars. Additionally, these instruments needed a professional-sized telescope to carry their weight. A version of the instrument called \mbox{Mini-HIPPI} \citep{Bailey2017} was made for smaller telescopes but suffered from the same PMT-related limitations.

One of the first good high precision polarimeters was developed by \citet{Kemp1981}, using photo-elastic modulators (PEMs) -- crystals that oscillated at tens of thousands of Hz; this technology requires a lock-in amplifier and is somewhat cumbersome as a result. Two other modern high-precision polarimeters (\mbox{PlanetPol}, \citealp{Hough2006}; POLISH/-2, \citealp{Wiktorowicz2015}) also used PEMs. Like HIPPI/-2, they could only be mounted on larger telescopes ($>$0.5-m).  Other high-precision schemes involved spreading the light out in a circular path on a CCD \citep{Clarke2002}, or defocussing the instrument for bright targets (\mbox{DIPol/-2/-UF}, \citealp{Piirola2014, Piirola2021}) to avoid detector saturation, limiting their effectiveness on the brightest targets. Aside from \mbox{PlanetPol} which had a precision better than 2 ppm (but had the additional drawback that it could only observe red and NIR wavelengths\footnote{Seeing noise is greater at shorter wavelengths.}, owing to its avalanche photo-diode detectors) the best precision achieved by these other instruments is 8-10 ppm.

The advent of CMOS sensors for astronomy (e.g. \citealp{Breacher2023}) has significantly advanced the state of the art, since now fast modulation can be achieved without appreciably sacrificing efficiency. The newly commissioned POPO is reported to achieve a precision of 5-10 ppm \citep{Takahashi2025}; it is the second to use CMOS technology for the detector -- pairing twin science grade CMOS cameras with an FLC modulator and dialectic beam-splitting cube analyzer\footnote{Notably, with this instrument \citet{Takahashi2025} investigated the difference in precision between a double- and single-beam configuration with fast modulation, finding the single-beam precision to be not far off the double-beam instrument.}. Like many of the instruments described above, it is too bulky for use on amateur-sized telescopes. Whereas, the original PICSARR instrument was designed specifically for that purpose.

\subsection{The PICSARR Polarimeter}
\label{sec:picsarr}

First commissioned in 2021, the Polarimeter using Imaging CMOS Sensor and Rotating Retarder (PICSARR; \citealp{Bailey2023}) made high precision stellar polarimetry feasible on small telescopes -- inspiring at least one copy \citep{Kamara2024}. Its 3D-printed plastic case, off-the-shelf optics and hardware, and inexpensive CMOS camera -- of the type marketed to amateur astrophotographers -- made it inexpensive and belied its high performance. It features a $\mathrm{MgF_2}$ Wollaston prism analyzer that directs the orthogonally polarized components of starlight onto the single detector -- which typically takes data at a rate of 83 fps (the sky signal from each of the two beams overlaps the other, cancelling the sky polarization -- but not the flux). In front of that in the light path is a ZWO electronic filter wheel and a super-achromatic half-waveplate modulator that is rotated by a belt-driven mechanism to provide fast modulation -- smoothly varying the polarization state measured by each beam.

Five PICSARR-class instruments are currently in operation, across four institutions, on telescopes with apertures ranging in size from 8 to 36 inches. They are being used to study, stars, planets (especially Solar system planets), the interstellar medium, and nebulae. Chief among their applications are programs to study variable stars, including eclipsing and non-eclipsing close binaries, $\beta$ Cep variables, $\alpha$ Cyg variables, Long Period Variables, Be stars and other eruptive variables. Most of these are still in progress, but there are papers on Deneb \citep{Cotton2024a}, and Venus \citep{Bailey2026}. The second of these also makes use of the imaging functionality of the instrument. 

Three of these polarimeters, that we denote PICSARR-2, incorporate a hollow shaft stepper motor to smoothly accomplish the rapid modulation needed for high precision. This is a significant improvement on the original belt drive system, which enables longer frame exposures for the observation of fainter targets, and smoother operation improved imaging and stability on bright targets. The purpose of the current paper is to describe the upgraded instrument, report its performance, and provide brief examples of the variable star research being done with it.

In Sec. \ref{sec:picsarr2} we first describe the instrument, focusing on changes from the original PICSARR. In Sec. \ref{sec:calibration} the methods of calibration are described in an instructional manner. Similarly, in Sec. \ref{sec:performance} we relate the results of performance characterization tests. In Sec. \ref{sec:sci_eg}, examples of current observing programs are given, and an example of the imaging capability of the instrument is shown in Sec. \ref{sec:NGC_7027}. A discussion and the conclusions are in Secs. \ref{sec:discussion} and \ref{sec:conclusions}, respectively.

\begin{figure*}[t!]
\includegraphics[trim={0.5cm, 6.2cm, 1cm, 3cm}, clip, width=\textwidth]{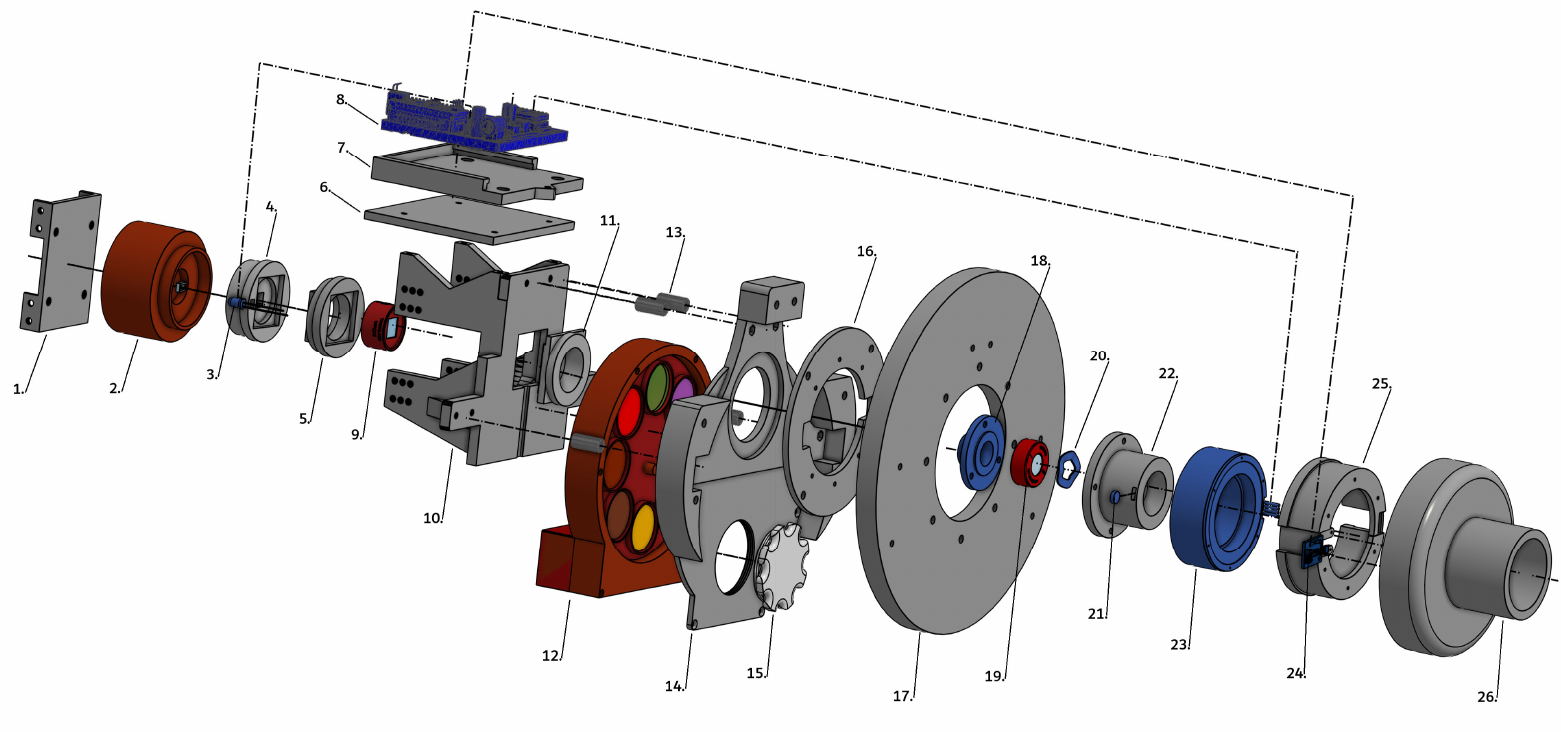}
\caption{An exploded diagram of the PICSARR-2 instrument (without any screws, nuts or bolts). Plastic parts are shown in grey, optical components in red, and other components in blue. The numbers indicate the individual parts, as follows: \mbox{1. ASI Camera support plate}, 2. Camera (ZWO ASI 462MM or ASI 662MM), 3. LED (for providing the reference signal), \mbox{4. Camera insert}, 5. Wollaston mount extension (optional for changing beam spacing)  6. Circuit board mounting plate, \mbox{7. Circuit} board tray (attached to 6. with double-sided tape), 8. Circuit board, 9. Wollaston prism (Thorlabs MgF$\mathrm{_2}$), 10. ASI Camera support, 11. Wollaston mount, 12. ZWO 8-slot Filter wheel for 1.25-inch filters (the cover is not used in the instrument; filters also shown), 13. Support posts ($\times$4), 14. Filter wheel interface, 15. Filter wheel plug (allows access to filters without removing instrument from telescope), 16. Motor mount plate, 17. Instrument mounting plate (representation only, see Fig. \ref{fig:P2}), 18. Waveplate holder top (the optical alignment of the instrument is adjusted by means of the three screws that attach this part), 19. Waveplate (Thorlabs AHWP05M-580), 20. Spring washer, 21. Magnet (for triggering Hall sensor, held in with glue), 22. Waveplate holder, 23. Hollow shaft stepper motor, 24. Hall sensor, 25. Motor mount, 26. 2-inch eyepiece adapter (optional; also called the nose piece). The dash-dot lines with vertical components show the electrical connexions.}
\label{fig:exploded}
\end{figure*}

\begin{figure*}[t!]
\includegraphics[width=0.498\textwidth]{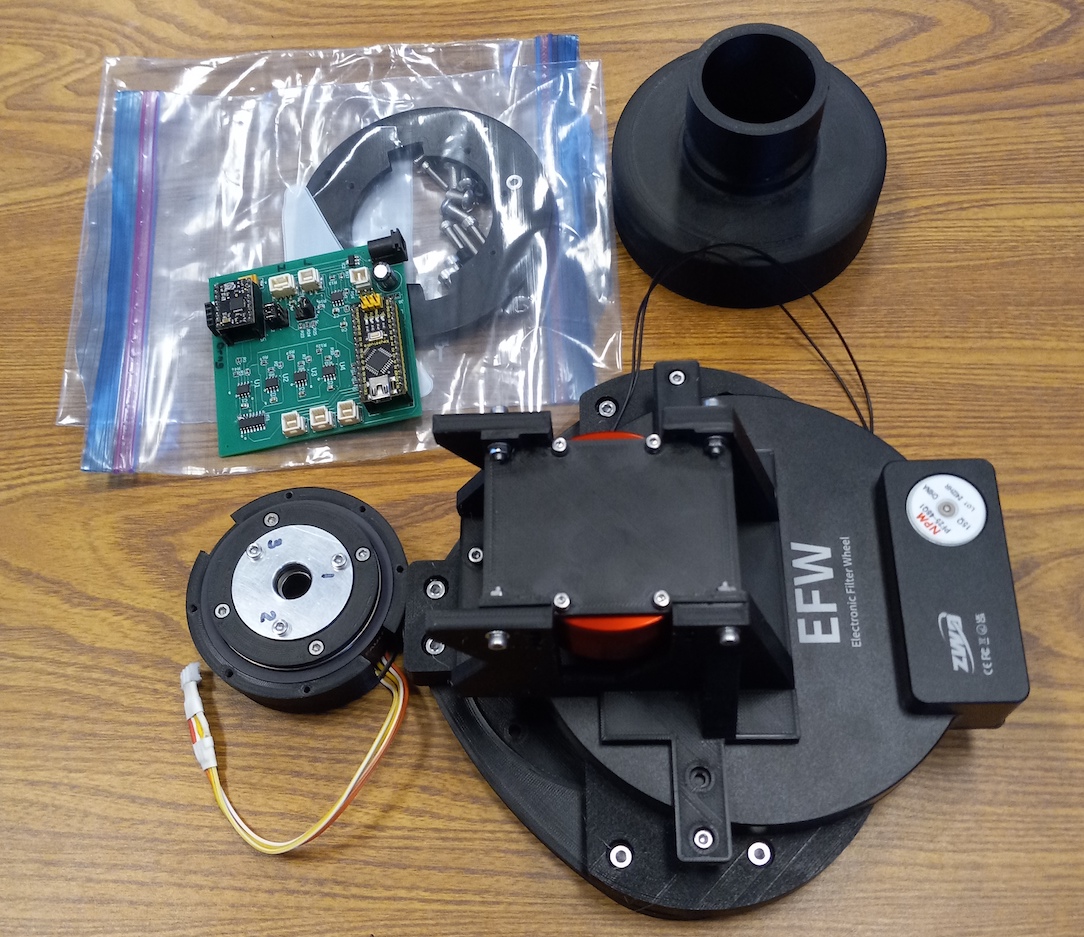}
\includegraphics[width=0.498\textwidth]{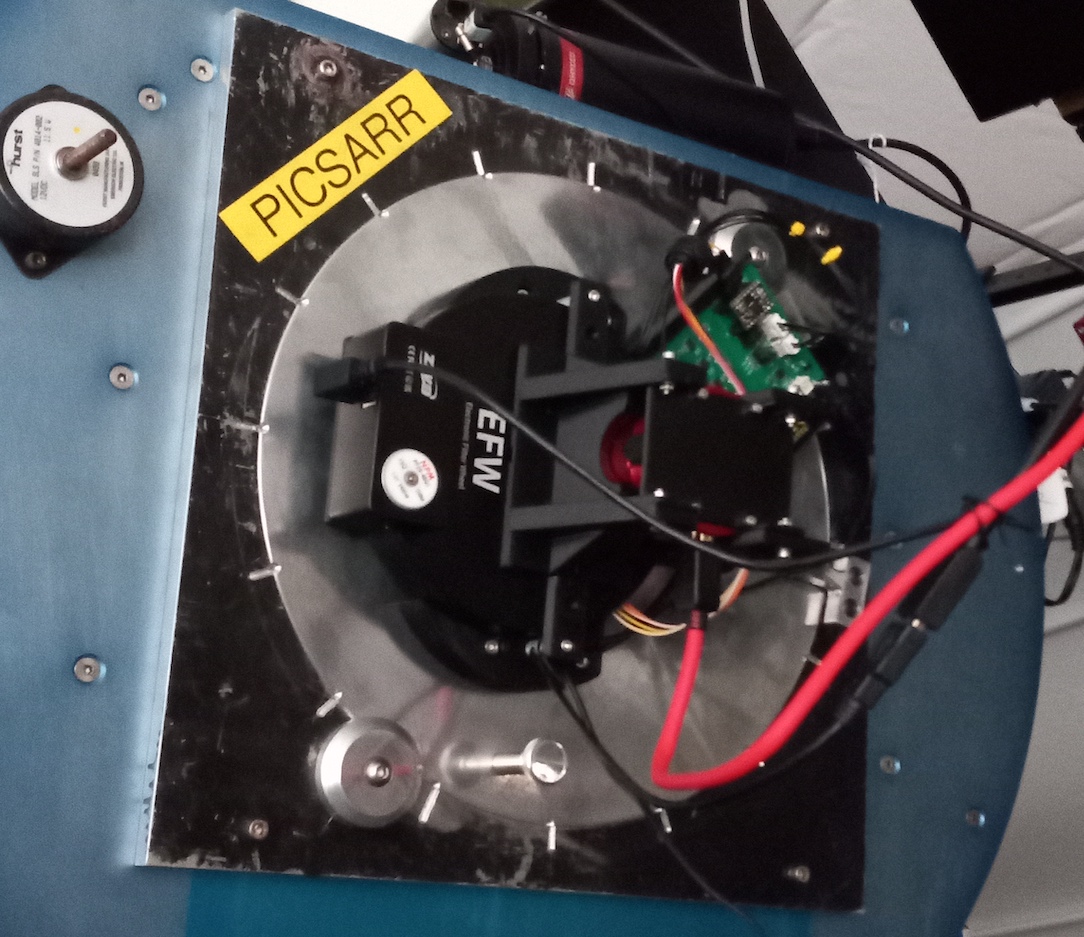}
\caption{Left: partially assembled PICSARR-2, with the electronics board top left, modulator assembly at bottom left, the camera assembly at bottom right, and the 2-inch eyepiece adapter for smaller telescopes on the top right. Right: PICSARR-2 attached to its black and silver aluminum mounting plate on MIRA’s 36-inch telescope GAP. In this configuration, the modulator assembly is within the GAP.}
\label{fig:P2}
\end{figure*}

\section{Methods}
\label{sec:methods}

\subsection{The PICSARR-2 Polarimeter}
\label{sec:picsarr2}

The three copies of PICSARR-2 currently in operation incorporate subtle differences in their componentry and design. The first two were built by one of us (JB), the third, to be described here, was fabricated at the Monterey Institute for Research in Astronomy (MIRA). It weighs $\sim$3.5 lbs and cost only \$5600 to build. The modular design of these instruments allows for differing modulators and detectors. MIRA’s instrument also includes minor layout changes to accommodate the short back-focal distance of the Cassegrain port of the Guidance and Acquisition Package (GAP, \citealp{Weaver1985b}) on the f/10 MIRA 36-inch telescope \citep{Weaver1985a}. This telescope, used for the majority of the observations reported here, is located at MIRA's Oliver Observing Station \citep{Perkins2022}, which has a high percentage of photometric nights \citep{Weaver1989}, and a typical seeing of 0.75 arcsec, excluding winter \citep{Hutter1997}.

\begin{figure*}[t!]
\includegraphics[width=\textwidth]{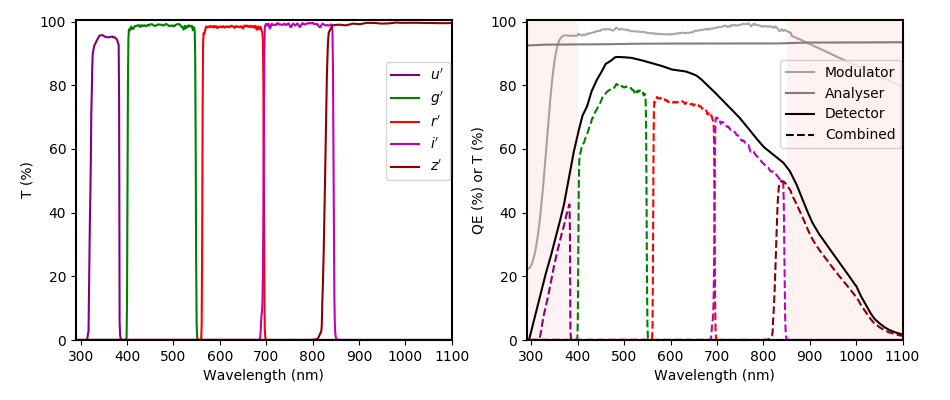}
\caption{Left: Chroma SDSS filter transmission as a function of wavelength. Right: the relative response of PICSARR-2 in each filter when taking account of the ASI 462MM camera QE and the transmission of the other optical components. Note: pink shaded regions are estimated from second order data.}
\label{fig:P2eff}
\end{figure*}

An exploded diagram of PICSARR-2 is shown in Fig. \ref{fig:exploded}, for comparison with figure 4 in \citet{Bailey2023} showing PICSARR. What distinguishes PICSARR-2 from PICSARR is the use of a Faulhaber DM66200H hollow shaft stepper motor to spin the half-waveplate modulator, which is housed within it. The motor features 200 steps of 1.8 degrees, but for smoother operation uses 1/16 micro-steps. This replaces the previous belt-drive system. Through the use of a custom electronics board, the motor can be manually stepped into one of 16 fixed positions corresponding to spacings of 22.5 degrees (equal to 45 degrees in position angle), or continuously rotated at rates corresponding to frame exposures of 12, 50, 200, or 500 ms (83, 20, 5, 2 fps, respectively); where there are 16 frames per complete waveplate rotation block (always processing data as complete blocks serves to cancel instrumental effects).

The modulator is held in its own cradle mounted within the motor; it is press fit against a spring washer. Three screws are used to adjust the tilt of the modulator within its cradle for better optical alignment; the adjustment plate these screws pass through was machined out of aluminum after the plastic version was found to be insufficiently rigid (see the left panel of Fig. \ref{fig:P2}). This mechanism has substantially reduced beam motion as a function of rotation angle compared to PICSARR.

On MIRA’s 36-inch telescope the instrument is attached to an aluminum mounting plate affixed directly to the Cassegrain face of the GAP. The mounting plate incorporates a turntable that allows adjustment of the position angle on the sky of the detector (Fig. \ref{fig:P2} right panel). The modulator assembly and the camera assembly, incorporating all the other optics, are attached to the mounting plate separately. In this way, the plate and the mass of the GAP act to damp vibrations induced by the motor in the modulator assembly.

Rotation of a half-waveplate is a common way to achieve fast modulation in polarimetry. What is often overlooked though are the mechanical considerations in doing this effectively. Vibration from the rotating parts can act like seeing noise in its effect on the images at the detector, and it increases in frequency as the modulation rate is increased. Vital considerations therefore are a motor with a smooth action, along with vibration damping.

For the observations reported here, PICSARR-2 was configured with a Thorlabs AHWP05M-580 achromatic half-waveplate modulator, a Thorlabs WPM10 $\mathrm{MgF_2}$ Wollaston prism analyzer with a beam separation of $1^\circ\,20'$, and made use of an SDSS $ugriz$ filter set by Chroma. Two runs were completed on the 36-inch telescope with two different cameras, the ZWO ASI 462MM and the ZWO ASI 662MM. The ASI 662MM has a larger dynamic range at the cost of a greater read noise compared to the ASI 462MM but otherwise these are very similar cameras. We designate the runs MOP2025JUL and MOP2025SEP, where the first two letters are an observatory/telescope code (for MIRA’s 36-inch telescope at its Oliver Observing Station), P stands for PICSARR, and the remainder refers to the month the instrument was first mounted. The MOP2025JUL run is split into an A and B sub-run as a result of a minor set-up change. A further run (MWP2024DEC) is underway on a small telescope, the details of which are saved for Sec. \ref{sec:small}.

Fig. \ref{fig:P2eff} shows the transmission of the filter set, the optical components, and the Quantum Efficiency (QE) of the ASI 462MM camera. The modulator batch was specially characterized by Thorlabs in the region 400 to 850 nm. Outside this range its performance has been estimated based on the data available from previous batches. Similarly, the camera QE is estimated outside of 400 to 1000 nm based on data from Sony on chips of the same generation.

The software reduction package, including the bandpass model, is largely unchanged from that described for PICSARR \citep{Bailey2023}.

\subsection{Calibration}
\label{sec:calibration}

Polarimeter characterization and calibration is very important for making reliable measurements that can be compared to the work of others. The basis of this is an accurate bandpass model as a function of wavelength. With PICSARR-2 we have mostly relied on manufacturer data for the transmission of each optical component and the detector quantum efficiency (QE) (Fig. \ref{fig:P2eff}), as well as the modulator’s modulation efficiency and fast-axis deviation. However, we always regard this data as approximate until verified on-sky against established standard stars and corrections applied where indicated.

For any broadband linear stellar polarimeter, there are two on-sky calibrations that must be performed for every new mounting of the instrument: measurement of the telescope polarization, and a position angle calibration. We perform these first before other characterization, and then iterate to establish the performance of the instrument.

\begin{figure*}
\includegraphics[width=\textwidth]{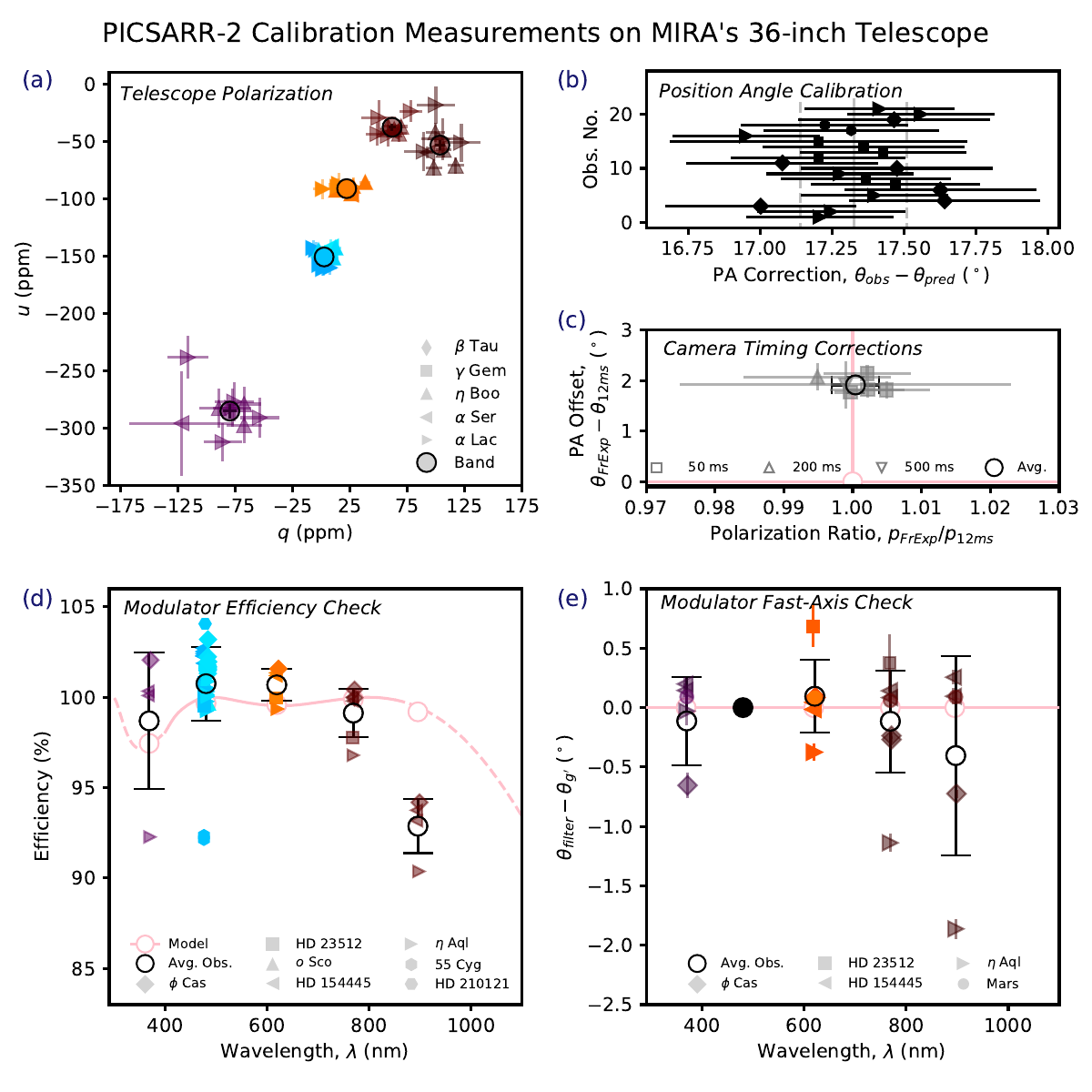}
\caption{PICSARR-2 calibration measurements on MIRA's 36-inch telescope. (a) Telescope polarization in $u^\prime$ (purple), $g^\prime$ (cyan/blue), $r^\prime$ (orange), $i^\prime$ (brown-red), and $z^\prime$ (brown-black) bands for run MOP2025JUL (the colors represent $\lambda_{\rm eff}$ of the passband), (b) position angle calibration for run MOP2025SEP, (c) camera timing corrections for the ASI 662MM, (d, e) moduator efficiency and fast-axis checks, respectively, for the Thorlabs AHWP05M-580. An explanation of the symbols can be found in the sections \ref{sec:tp}, \ref{sec:pa}, \ref{sec:cam_t_corr}, \ref{sec:mod_eff_chk}, and \ref{sec:mod_fa_chk}.}
\label{fig:calibration}
\end{figure*}

\subsubsection{Telescope Polarization}
\label{sec:tp}

The first calibration is the telescope polarization (TP) measurement, which defines the zero point of Stokes $q$ and $u$. It needs to be determined independently for each passband used. The simplest method is to take the straight average of several observations of low polarization standard stars, the result of which is subtracted from all the other observations in the same passband. Fig. \ref{fig:calibration}(a) shows the TP data for our MOP2025JUL run; the standards used are listed in the key, these are assumed to have no polarization. In reality these stars have small polarizations, the lack of knowledge of which places a limit on the accuracy obtainable.

In Fig. \ref{fig:calibration}(a) the TP data is shown in the Equatorial frame (See Sec. \ref{sec:linear_pol}). However, if the TP is significant, one should first make these calculations in the instrument frame and subtract the results before a position angle calibration (see Sec. \ref{sec:pa}). The slope of $\mathrm{TP_{[q/u]}}$ with wavelength limits the accuracy of this method. If it is large compared to the needed accuracy, then a second order correction for the color of the stars (standards and science targets) may be needed. Here, the implied TP difference between blue and red stars is smaller than the nominal errors for a typical observation of the brightest stars (see Sec. \ref{sec:pol_uncertainty}).

\subsubsection{Position Angle Calibration}
\label{sec:pa}

The instrumental $Q$ and $U$ axes are a function of the layout of instrument hardware components.  The position angle calibration rotates the $Q$-$U$ frame from that of the instrument to the Equatorial frame (See Sec. \ref{sec:linear_pol}). This is accomplished by a comparison of the observed position angle, $\theta$, of high polarization standard stars to their predicted values -- determined from literature parameters. The simplest method would take as a prediction previously observed values in the same band and take a straight average of all observations. We use a more sophisticated method explained fully in \citet{Cotton2024b}, that takes account of wavelength and time dependence, and weights each measurement according to the RMS sum of the measured error, the known variability in the standard, and an instrument-telescope precision, $\eta(\epsilon_{i\theta})$, which is the median of error-corrected standard deviations in measurements of the same stars during the same run with the same instrumental set-up. From the MOP2025JUL A and B runs, we calculated this parameter to be 0.0845 degrees – about half the adopted value for HIPPI-2 on the same telescope.

Fig. \ref{fig:calibration}(b) depicts graphically the position angle calibration for the MOP2025SEP run. The symbols corresponding the standards used can be found in the key of Fig. \ref{fig:calibration}(d). The vertical grey line shows the adopted value, and the dashed lines $\pm$ the weighted standard deviation, 0.185 degrees, whereas the nominal RMS error in the determination is 0.0681 degrees. These values are comparable to the potential error that might be introduced if a TP subtraction was not carried out first. For any given standard this could be estimated, assuming the TP offset is $p_e$, as \begin{equation}\theta_e=28.65\left(\frac{p_e}{p}\right),\end{equation} in degrees, where $p_e$ is the error in polarization and $p$ the standard’s nominal polarization -- typically several percent.

\subsubsection{Camera Timing Corrections}
\label{sec:cam_t_corr}

The cameras used with PICSARR-2 employ a rolling shutter: each row of pixels is read out in sequence. To achieve the fastest modulation speeds the modulator is rotated continuously, resulting in a slight rotation between rows. We correct for this using a formula for frame rotation determined empirically from repeat standard observations made previously with the original PICSARR using the ASI 462MM: \begin{equation}\theta_{row}=\frac{0.6252}{FrExp} b\left(\frac{h}{2}-y\right),\end{equation} where $b$ is the pixel binning of the camera (i.e 2 for $\mathrm{2\times2}$), h the window height in pixels, y the average row center of the star for the video sequence, and $FrExp$ the frame exposure in ms. For the ASI 662MM an adjustment is made to account for the maximum frame rate difference. 

The rolling shutter correction is less important with \mbox{PICSARR-2} than PICSARR because the optics are better aligned resulting in less circular motion of the star on the detector. Additionally, we are now able to employ autoguiding directly from the \textsc{SharpCap} (www.sharpcap.co.uk) display using the TCS \textsc{Helper} program written by one of us (LB; www.mira.org/software.htm), which keeps the star images much closer to the target pixel. 

Because the modulator is driven continuously at a high rate, the row readout can lead to a modulation efficiency loss or position angle rotation when the readout time is a significant fraction of the total. In Fig. \ref{fig:calibration}(c), we plot data obtained from sequential measurements made of the same standards with different frame rates for the ASI 662MM camera. By the polarization ratio, $\frac{p_{\mathrm{FrExp}}}{p_{\mathrm{12ms}}}$, measured, these show no significant efficiency loss. However, there is a small position angle offset of 1.913 degrees for the fastest 12 ms frame rate.

\subsubsection{Modulator Efficiency Check}
\label{sec:mod_eff_chk}

To check the accuracy of the lab data provided on the Thorlabs AHWP05M-580 modulator efficiency, in Fig. \ref{fig:calibration}(d) we plot the measured polarization of standard stars in different passbands as a percentage of the predicted value (derived from \citealp{Cotton2024b}), with the modulator efficiency correction removed. The modelled efficiency is shown by the pink line. It is clear that there is a significant scatter in the literature polarizations of the standards. Nevertheless, the $z^\prime$ band data is significantly different from the prediction – which had to be estimated from prior modulator batch data in the corresponding wavelength range – and so we scale this band by a factor of 1.0721 to compensate. We could scale the other bands as well to get better agreement, but it is not certain that this would lead to greater accuracy.

\subsubsection{Modulator Fast-Axis Check}
\label{sec:mod_fa_chk}

A half-waveplate rotates the plane of polarization by flipping the polarization direction around the fast-axis. To modulate the polarization, PICSARR-2 rotates a half waveplate, continuously changing the angle of the fast-axis relative to the incoming light. In some modulator designs, the angle of the fast-axis can vary with wavelength. For an accurate measurement this behavior must be characterized and included in a bandpass model.

The AHWP05M-580 is an achromatic half-waveplate. Therefore, its fast-axis should not change as a function of wavelength. In PICSARR, a super-achromatic version was used, which did vary quite a lot, requiring a more complex bandpass model for corrections. (Thorlabs generously recharacterized a SAHWP05M-700 modulator for us after initial observations revealed a significant discrepancy with the nominal lab data.) It may also have contributed to position angle imprecision through chromatic aberration apparent at high airmass. 

In Fig. \ref{fig:calibration}(e) we check that there really is no fast-axis deviation with wavelength. Observations of the same standard in different bandpasses are made in sequence, and the difference in position angle between the $g^\prime$ band and the other filters are calculated (data from \citealp{Cotton2024b}). We also included an observation set of the planet Mars as an additional standard – to first order Solar system bodies have polarizations that are either parallel or perpendicular to the Sun-Target-Observer plane, regardless of wavelength. The error bars show the standard deviations for each passband – there are no significant deviations. In fact, our observations serve to highlight that the wavelength dependence of $\theta$ for some standards is poorly characterized. \mbox{HD 154445} looks to be the best standard, with deviations comparable to the Martian data.

We note in passing, that planetary data can serve as an alternative to high polarization standard stars for position angle calibration. If using this method the observer should carefully calculate the angle of the scattering plane relative to the equatorial frame at the time of the observation. One should also keep in mind that the polarization will periodically flip between plane parallel and perpendicular, and that this does not occur at the same phase for every passband.

\subsection{Performance Characterization}
\label{sec:performance}

\begin{figure*}
\includegraphics[width=\textwidth]{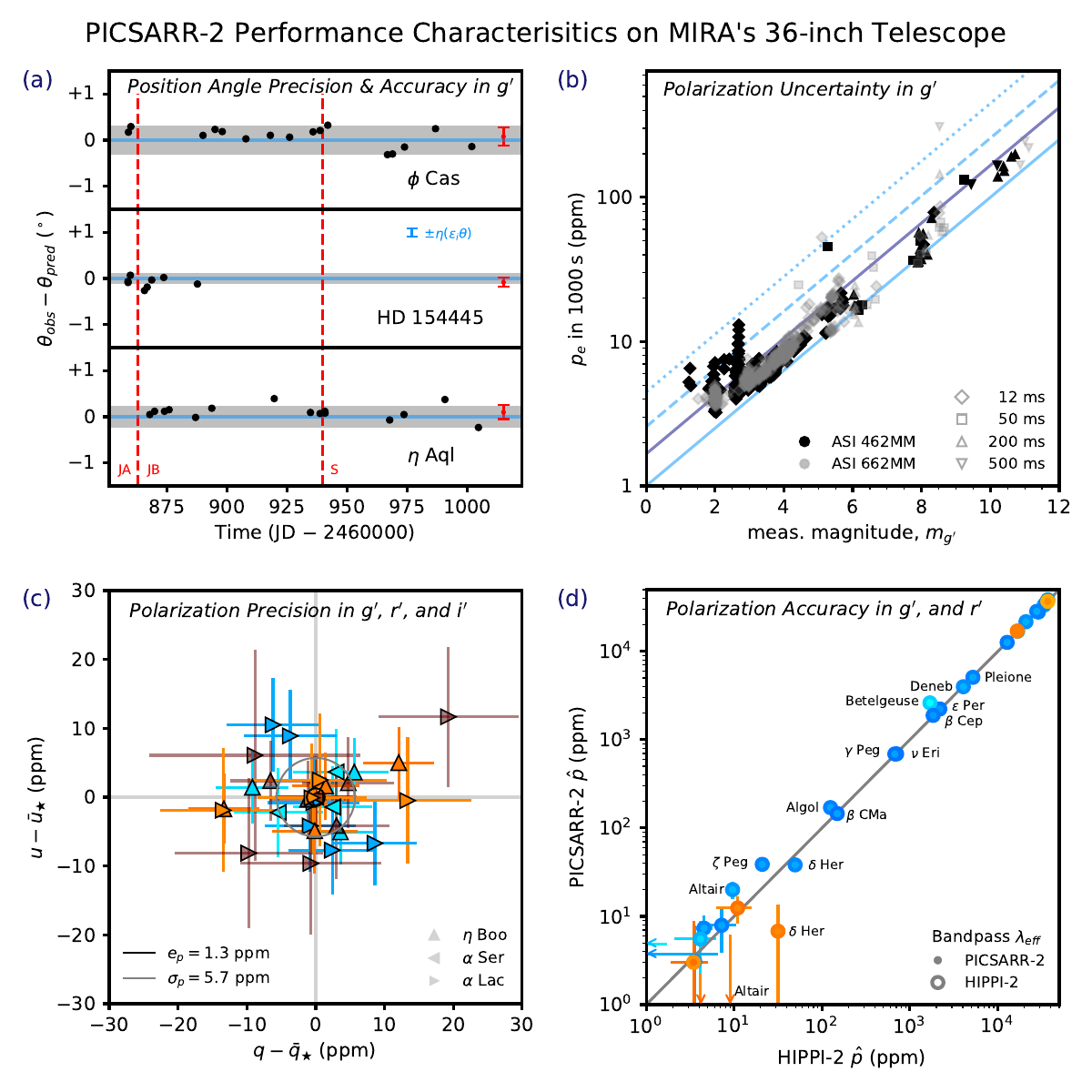}
\caption{Performance characteristics of PICSARR-2 on MIRA's 36-inch telescope. (a) Position angle accuracy and precision in $g^\prime$, (b) polarization uncertainty in $g^\prime$, (c) polarization precision in $g^\prime$ (blue/cyan), $r^\prime$ (orange) and $i^\prime$ (brown), where the exact colors correspond to $\lambda_{\rm eff}$, (d) polarization accuracy in $g^\prime$ and $r^\prime$. An explanation of the symbols can be found in the text in Sections \ref{sec:pa_prec_acc}, \ref{sec:pol_uncertainty}, \ref{sec:pol_prec}, and \ref{sec:pol_acc} respectively. In panels (c) and (d) the colors reflect the effective wavelengths of the passbands, and account for star color.}
\label{fig:performance}
\end{figure*}

\subsubsection{Position Angle Precision \& Accuracy}
\label{sec:pa_prec_acc}

Position angle precision and accuracy are important, not just for observations of large polarization variables, but also for small polarization variables at distances large enough to have significant interstellar polarization. The data from different observing runs can only be combined reliably, regardless of the nominal errors on the individual observations, if the telescope position angle is accurately established.

Figure \ref{fig:performance}(a) presents repeat observations of three high polarization standards, across all three sub-/runs (delineated by the dashed red lines), as differences from the model predicted position angle -- indicated by the horizontal blue lines. The height of the grey backing around the blue lines represents the established variability of each standard across a 10-year period \citep{Cotton2024b}. The number of observations falling beyond the grey area is fewer than expected for a normal distribution. The red data points and error bars at the right of frame show the mean observation and standard distribution; this is a formal match for expectations in all three cases. 

\subsubsection{Polarization Uncertainty}
\label{sec:pol_uncertainty}

Figure \ref{fig:performance}(b) presents the errors for all of the stellar data acquired in $g^\prime$ with PICSARR-2 on MIRA’s 36-inch telescope. These are plotted against the measured magnitude of the star offset such that the median measured magnitude matches the median literature derived value in $g^\prime$ from our bandpass model. The nominal errors are determined statistically from the standard deviation of each individual measurement of $q$ and $u$ from the 16-frame blocks that make up a video sequence. Most, but not all, observations are 16 minutes (4$\mathrm{\times}$4 min $=$ 960 s). For convenience, all the errors have been scaled to a 1000 s equivalent exposure using \begin{equation}e_2=e_1 \sqrt{\frac{t_1}{t_2}}.\end{equation} Upward outliers are the result of rejected blocks which are still counted in the total exposure time. Blocks can be rejected when the signal is too low (e.g. clouds), too high (a single saturated pixel), when the Ordinary (O) star image is too far from the target pixel or the O and Extraordinary (E) beam spacing is anomalous (from e.g. poor seeing or tracking errors), or due to time sequence errors.

The performance of the two cameras are plotted in different shades in Figure \ref{fig:performance}(b). There are more upward outliers for the ASI 462MM camera for bright stars. Many of these points correspond to blocks rejected by saturation, often in variable seeing conditions. The larger full well depth of the ASI 662MM (38.2 ke$^-$ vs. 11.2 ke$^-$ for the ASI 462MM) reduces such instances. However, this comes at the cost of a larger read noise, and so the ultimate performance is not quite as good (black points below grey). At Gain 0 the read noise is 7 e$^-$ for the ASI 662MM and 2.5 e$^-$ for the ASI 462MM, and decreases quickly with gain. A High Contrast Gain (HCG) mode, which significantly lowers the read noise doesn’t kick-in until a much higher gain for the ASI 662MM; for specific values at other gains see ZWO’s documentation.

Neither trend in camera data is entirely smooth. Where longer frame exposures are chosen, read noise is reduced, but faster frame rates better reduce the effects of seeing noise, so there is a trade off, and this setting, along with the Gain, is chosen by the observer at the time of the observation based on the prevailing conditions. Around $m=5$, the slower 50 ms frame exposure is preferred (square symbols), and this corresponds to a slight increase in nominal error. Another step is noticeable around $m=8$, where the 200 ms frame exposure (up triangle) is necessary. This step is larger, and lends experimental support to \citet{Tinbergen1973}'s statement that at least 10 Hz is required to achieve high precision, even in excellent seeing conditions.

The solid blue diagonal line represents a purely photon-limited error increase for a system with 1 ppm error at $m=0$; the instrument’s best performance matches this for magnitudes about 3 to 8.5 but begins to deviate for both fainter and brighter stars, as expected. The same bright star deviation was seen with HIPPI-2, where it was mainly attributed to centering imprecision \citep{Bailey2020, Cotton2022b}, but here read noise is probably a larger contributor. The solid plum line is the equivalent photon-limited trend offset to correspond with the performance of HIPPI-2 on the same sized telescope (based on \citealp{Bailey2020}). In terms of exposure time, PICSARR-2 is performing about twice as well, which is attributable to the relative detector efficiencies. However, if one considers \mbox{HIPPI-2’s} observing overheads, the performance is almost four times better \citep{Cotton2022b}.

For comparison purposes, a dashed and dotted blue diagonal lines are drawn in Fig. \ref{fig:performance}(b) that scales the solid line for telescopes with apertures of 14 inches, and 8 inches (20 cm) -- the smallest telescope used with PICSARR-2.

\subsubsection{Polarization Precision}
\label{sec:pol_prec}

The instrument’s precision is established by making repeat observations of bright low polarization stars across multiple nights. For comparison with \citet{Bailey2023} we have combined data sets from the $g^\prime$, $r^\prime$ and $i^\prime$ bands in Fig. \ref{fig:performance}(c). The average standard deviation of $q$ and $u$ for each set plotted in this figure is $\sigma_p = 5.7$ ppm. This puts the instrument at least on par with the reported precisions of POLISH \citep{Wiktorowicz2015}, and POPO \citep{Takahashi2025}, and not far off of HIPPI \citep{Bailey2015} and \mbox{HIPPI-2} \citep{Bailey2020}; all commissioned on much larger telescopes. PICSARR-2's $\sigma_p$ is comparable to the individual observation errors; this figure would likely improve with longer observations. The metric $e_p$ seeks to establish the scatter independent of the nominal errors by a root-mean-square subtraction of the average error of each set (see \citealp{Bailey2020}) -- this is a check of whether the nominal errors are really representative under these conditions; it represents a limit on the achievable uncertainty for a single observation. We calculate $e_p = 1.3$ ppm for these observations. This is an improvement on both the figures of 11 and 17 ppm for PICSARR on smaller telescopes (14-inch and 8-inch respectively; \citealp{Bailey2023}) and \mbox{HIPPI-2} on the same telescope (1.7 ppm; \citealp{Cotton2022b}), which makes \mbox{PICSARR-2} the stellar polarimeter with the best reported precision by this metric. It is also better than the 1.6 ppm we calculate for PlanetPol based on comparable observations reported by \citet{Bailey2008}.

\subsubsection{Polarization Accuracy}
\label{sec:pol_acc}

One measure of the instrument’s polarization accuracy was already presented in Fig. \ref{fig:calibration}(d) as the agreement between the modelled and measured modulator efficiency. Yet, a precise determination was difficult because of the spread in the results for different standards. We contend this is a consequence of errors in the literature values rather than imprecision of the instrument. To test this, in Fig. \ref{fig:performance}(d) we plot mean PICSARR-2 observations against those in common with HIPPI-2 in the same bands ($g^\prime$ and $r^\prime$). The data has been debiased by the RMS subtraction of the errors as: 
\begin{equation}\hat{p} = \left\{\begin{matrix} \sqrt{p^{2}-p_{e}^{2}}, \hspace{1cm} p\geq p_{e}
 \\ \hspace{1.3cm} 0, \hspace{1cm} p<p_{e}
\end{matrix}\right.\end{equation} 
The plotted grey line has a slope of 1; the slope of a line fit to the points is 1.0051 $\pm$ 0.0001. The r$^2$ value is 0.99994. Together with the small error in the slope, this demonstrates a very consistent relationship. This agreement is good but indicates that PICSARR-2 is measuring slightly higher polarizations. This is to be expected, given the data presented in Fig. \ref{fig:calibration}(d), since the modulator efficiency of HIPPI-2 is determined by fitting to the standard literature values, which are slightly high compared to our bandpass model prediction.

In Fig. \ref{fig:performance}(d) we have labelled all of the stars not either low polarization or high polarization standards (and Altair which we have sometimes used as a standard). The log-log scale emphasizes lower polarization objects -- the distribution of these emphasises how their use in TP determination limits accuracy. The next most apparent outliers are the known polarimetric variables Betelgeuse \citep{Schwarz1986} and Algol \citep{Kemp1983}, for which we only have one or two observations with one or both instruments.

\subsubsection{Small Telescope Performance}
\label{sec:small}

\begin{figure*}
\includegraphics[width=\textwidth]{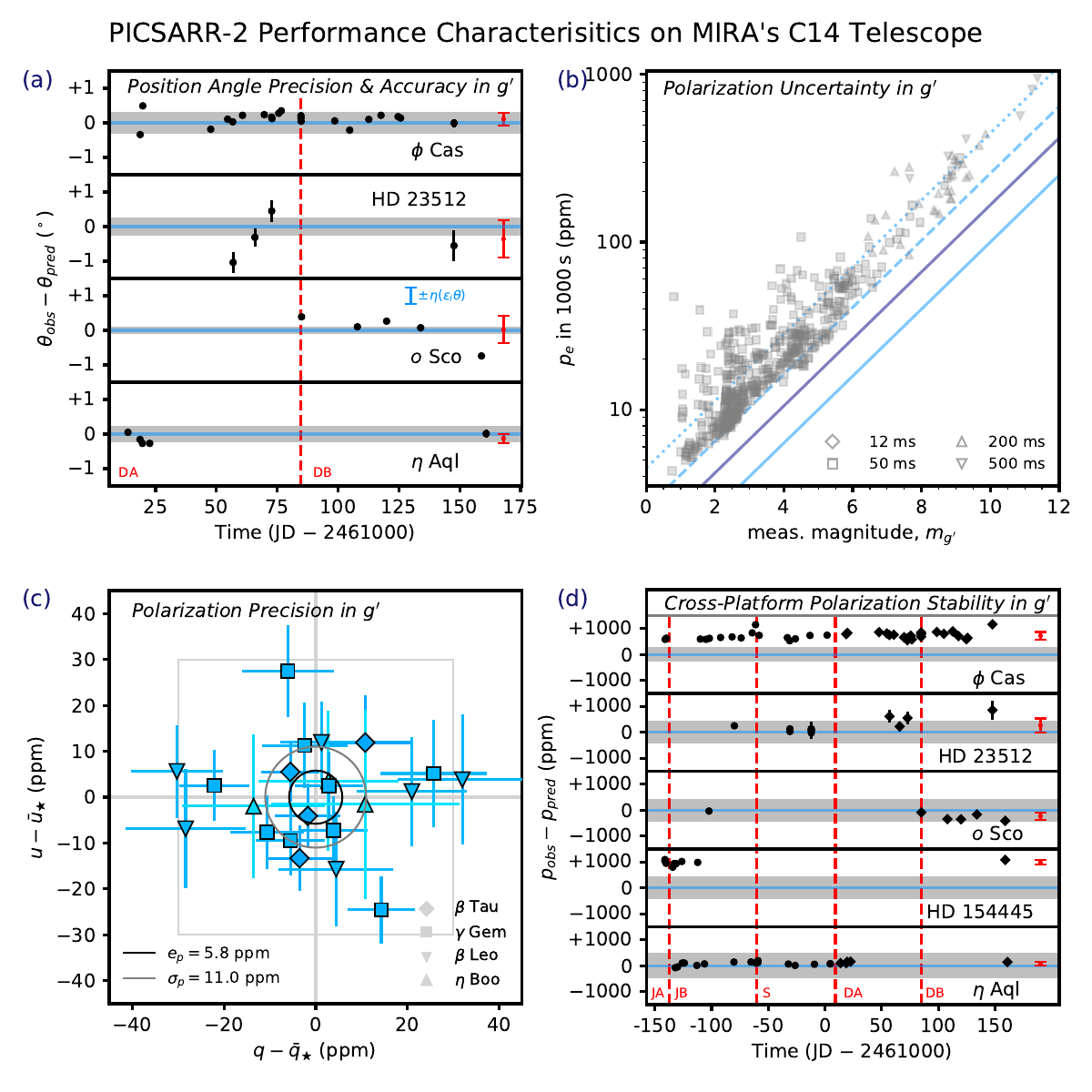}
\caption{Performance characteristics of PICSARR-2 on MIRA's C14 telescope. (a) Position angle accuracy and precision in $g^\prime$, (b) polarization uncertainty in $g^\prime$ for the ASI 662MM camera, (c) polarization precision in $g^\prime$: the colors reflect the effective wavelengths of the passband, and account for star color, (d) time series of high polarization standards observed with both telescopes demonstrating cross-platform polarization stability; the average predicted polarizations (thin blue lines) for each star are: \mbox{$\phi$ Cas} $=$ 32827 ppm, \mbox{HD 23512} $=$ 21534 ppm, \mbox{$o$ Sco} $=$ 38401 ppm, \mbox{HD 154445} $=$ 35146 ppm, \mbox{$\eta$ Aql} $=$ 16877 ppm. An explanation of the symbols can be found in the text in Sections \ref{sec:pa_prec_acc}, \ref{sec:pol_uncertainty}, and \ref{sec:pol_prec}. The square in panel (c) corresponds to the extent of the panel in Fig. \ref{fig:performance}(c) for easy comparison.}
\label{fig:wso_performance}
\end{figure*}

Subsequent to the tests described above, the same PICSARR-2 instrument was moved to the f/11 Celestron 14-inch (C14) telescope at MIRA's Weaver Student Observatory (WSO, \citealp{Babcock2008}), where it has operated for five months; the MWP2025DEC run has been split into an A and B subrun, where the division corresponds to a slight mounting adjustment. The WSO is at sea level, less than 1 mi from Monterey Bay and within a growing urban area, with all the challenges this implies. The instrument is mounted on the C14 using the 2-inch eyepiece adapter shown in Fig. \ref{fig:P2}. The ASI 662MM camera was used, mostly with 50 ms frame exposures, in compensation for the lower flux. Plots demonstrating the instrument's performance on this platform are shown in Fig. \ref{fig:wso_performance}.

Fig. \ref{fig:wso_performance}(a) shows position angle time series for standards observed more than once. The error bars represent only the nominal errors and do not account for any stellar variability (which is instead represented by the grey regions) or additional set-up imprecision. There is one clear outlier for $o$ Sco, but otherwise, the data matches what is expected well. From these data, we calculate $\eta(\epsilon_{i\theta})=$ 0.225 degrees for the telescope/instrument combination, based on the observations of $\phi$ Cas, $\eta$ Aql, HD 23512, and $o$ Sco. If the noted outlier is removed, we instead get 0.151 degrees. The difference to the value obtained on the 36-inch telescope will be partly due to polar alignment error, the effects of which worsen with declination. The position angle offset introduced by a polar misalignment can be calculated as 
\begin{equation}\Delta\theta=M_A \frac{\cos{H}\cos{L}}{\cos{\delta}}+M_E\frac{\sin{H}}{\cos{\delta}},\end{equation} where $L$ is the observer's latitude, $H$ is the hour angle of the observation, and $\delta$ the target's declination, all expressed in degrees. And $M_A$ and $M_E$ are the misalignments in azimuth and elevation, respectively.

Software Bisque's \textsc{TPoint} program reports $M_A=-$53.1 arcsec and $M_E=-$194.3 arcsec for the C14's current alignment and pointing solution. At present this is not accounted for in our calculations. Our most northern standard is $\phi$ Cas with $\delta=+58\:13\:53.8$ (J2000). So that the maximum deviation is a little more than 0.1 degrees.

Considering polarization uncertainty as a function of stellar magnitude in Fig. \ref{fig:wso_performance}(b), there is more scatter in the errors recorded than in Fig. \ref{fig:performance}(b). We attribute this to the conditions at each site. The OOS has some of the clearest skies in the country, whereas the WSO is situated in an area known for coastal fog \citep{Peterson1975, Kim2022}. The transparency is frequently degraded, especially at high airmass -- something not accounted for by our static bandpass model. Despite this, the best data points lie closer to the dashed line in Fig. \ref{fig:wso_performance}(b) than they do the solid line in \ref{fig:performance}(b), indicating the instrument is outperforming a purely photometric scaling of the 36-inch result. This supports the conclusion that read noise is a limiting contributor to performance.

The polarization precision is calculated as $\sigma_p=$ 11.0 ppm, and $e_p=$ 5.8 ppm from repeat 16 minute observations in $g^\prime$ of $\beta$ Tau, $\gamma$ Gem, $\beta$ Leo, and $\eta$ Boo; this is depicted graphically in Fig. \ref{fig:wso_performance}(c). As the natural seeing at the WSO is not as good, this in combination with the slower modulation might account for the precision not quite reaching the same level as on the 36-inch. There may be other reasons, but it hardly matters because better precisions are virtually unusable when, as here, other sources of noise -- like photon shot noise -- are more limiting.

The aggregate polarization accuracy on the C14 is no different to the 36-inch result. So, in Fig. \ref{fig:wso_performance}(d) we have instead plotted time series for high polarization standards observed with both telescopes as a way to gauge the cross-platform stability of the instrument. The thin blue lines mark $p$ calculated from literature data; it can be seen that both $\phi$ Cas and HD 154445 have higher polarizations than expected, and $\phi$ Cas is perhaps a bit more variable than reported in \citet{Cotton2024b}, while $\eta$ Aql\footnote{Readers may be surprised to see the first discovered Cepheid \mbox{\citep{Pigott1785}} used as a standard, but Cepheids pulsate only radially, which classically results in no change in polarization.} looks more stable (previously determined variability is represented by the grey shaded region). However, there is no significant difference between the 36-inch subruns (MOP2025JULA, MOP2025JULB, MOP2025SEP) and the C14 subruns (MWP2025DECA, MWP2025DECB), nor any indication of a performance shift in the 10 months of operation.

\citet{Cotton2024b} also found disagreement with the literature values for polarization magnitude in some standards, but eschewed recommendations for change because the combination of different HIPPI-class instruments made the $p$ precision less reliable than that for $\theta$. The stability of PICSARR-2 across telescopes and using different cameras offers a more favorable circumstance for investigating this. The original PICSARR instrument also performed well in this regard \citep{Bailey2023}, so combining the data for a better long term study is feasible.

\section{Science Examples}
\label{sec:sci_eg}

\citet{Bailey2023} provided some examples of the type of studies that PICSARR facilitates. They give a number of examples of what can be accomplished in Solar system science. A polarization map of the Homunculus Nebula is presented as a further example of the imaging capability of the instrument. An example of stellar science was given in the form of multi-band observations of the $\mu^1$ Sco binary system, where the variability is due to photospheric reflection (also see \citealp{Cotton2020}). Below we add a broad array of examples looking at stellar variability.  

\begin{figure*}
\centering\includegraphics[width=0.79\textwidth]{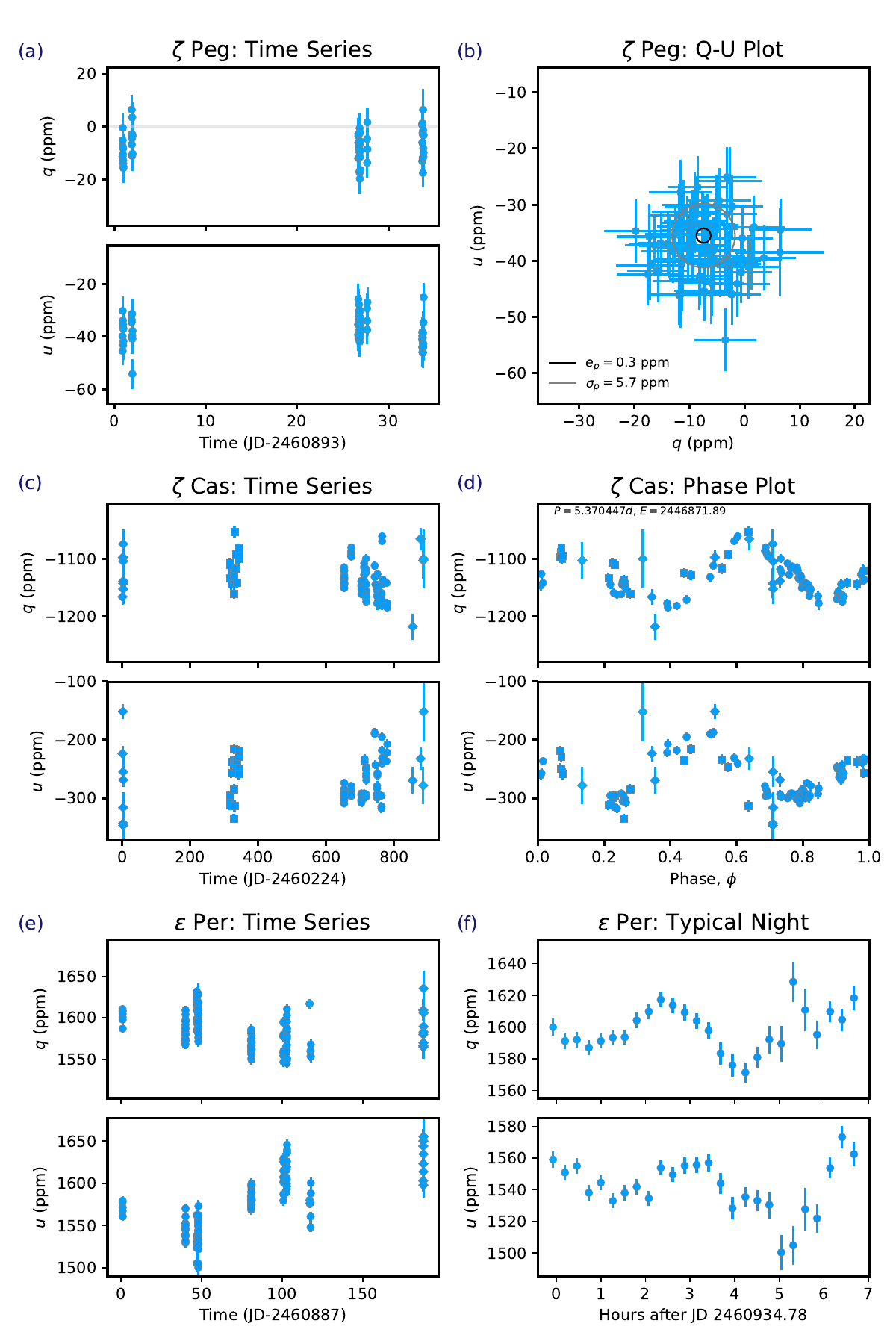}
\caption{Example data from the pulsating star program. \mbox{PICSARR-2} data from MIRA's 36-inch telescope is shown as circles, earlier PICSARR data from the same telescope as squares, and PICSARR/-2 data from other smaller telescopes as diamonds. All observations were made in the $g^\prime$ filter.}
\label{fig:pulsating}
\end{figure*}

\subsection{Pulsating Stars}
\label{sec:pulsating}

The main observing program for PICSARR-2 on MIRA’s 36-inch telescope aims to do polarimetric asteroseismology on bright $\mathrm\beta$ Cep and Slowly Pulsating B-type (SPB) variables. This pushes the performance of the instrument to the limit. Many hundreds of observations per star are required to detect pulsations with amplitudes expected to be $\approx20$ ppm or less \citep{Cotton2022a}. A Fourier analysis is required to identify different pulsation periods. Then the data has to be combined with high precision photometry and high-resolution spectroscopy to make mode assignments (in terms of $\ell$ and $m$), which ultimately allows modelling of the interior of the star. The focus is on B-type stars because traditional asteroseismology techniques alone are unable to identify the pulsation modes, and polarimetry offers complimentary data. Fig. \ref{fig:pulsating} shows example data for three $\sim$3rd magnitude stars in the program.

\subsubsection{\texorpdfstring{$\mathrm{\zeta}$}{zet} Pegasi}
\label{sec:zet_Peg}

One of the stars, $\mathrm{\zeta}$ Peg, exhibits no significant polarization variability. The $Q$-$U$ plot in Fig. \ref{fig:pulsating}(b) includes 64 $g^\prime$ data points from the MOP2025JUL run, displaying less scatter than the standards used to estimate the instrument’s precision in Fig. \ref{fig:performance}(c). This is an SPB star with a single low amplitude ($\Delta m \approx 0.001$) photometric mode with a period of 22.952 $\pm$ 0.804 h \citep{Goebel2007}. We do not expect a detectable signal from it, so the result provides more evidence of the instrument’s capability.

\subsubsection{\texorpdfstring{$\mathrm{\zeta}$}{zet} Cassiopeia}
\label{sec:zet_Cas}

The second example is $\mathrm{\zeta}$ Cas. \citet{Neiner2003} found a non-radial pulsation mode with $\ell = 2 \pm 1$ at $P = 1.57$ d, so it is a good candidate for a significant polarimetric signal. However, the star also has a weak dipole magnetic field ranging from about $-$150 to $+$150 G \citep{Briquet2016}. Modulated by its 5.37 d rotational period this produces a characteristic double peaked phase curve -- Fig. \ref{fig:pulsating}(d) – with an amplitude of $\approx$200 ppm. This signal will have to be removed to search for the signature of pulsation.

\subsubsection{\texorpdfstring{$\mathrm{\epsilon}$}{eps} Persei}
\label{sec:eps_Per}

The final example, $\mathrm{\epsilon}$ Per, shows clear variability. In the single night example shown in Fig. \ref{fig:pulsating}(f) there is a signal with up to 100 ppm amplitude. This star is a hybrid pulsator with four dominating photometric periods of 4.47, 3.84, 3.04 and 2.26 h, which are thought likely to be non-radial modes with $m = -$3, $-$4, $-$5 and $-$6 respectively \citep{Gies1988}. Any or all of these modes might produce a large sinusoidal polarimetric signal of the type seen in Fig. \ref{fig:pulsating}(f), and beating between them is likely responsible for the larger range of the time series in Fig. \ref{fig:pulsating}(e).

\subsection{Other Stars}
\label{sec:other_stars}

\begin{figure*}
\includegraphics[width=\textwidth]{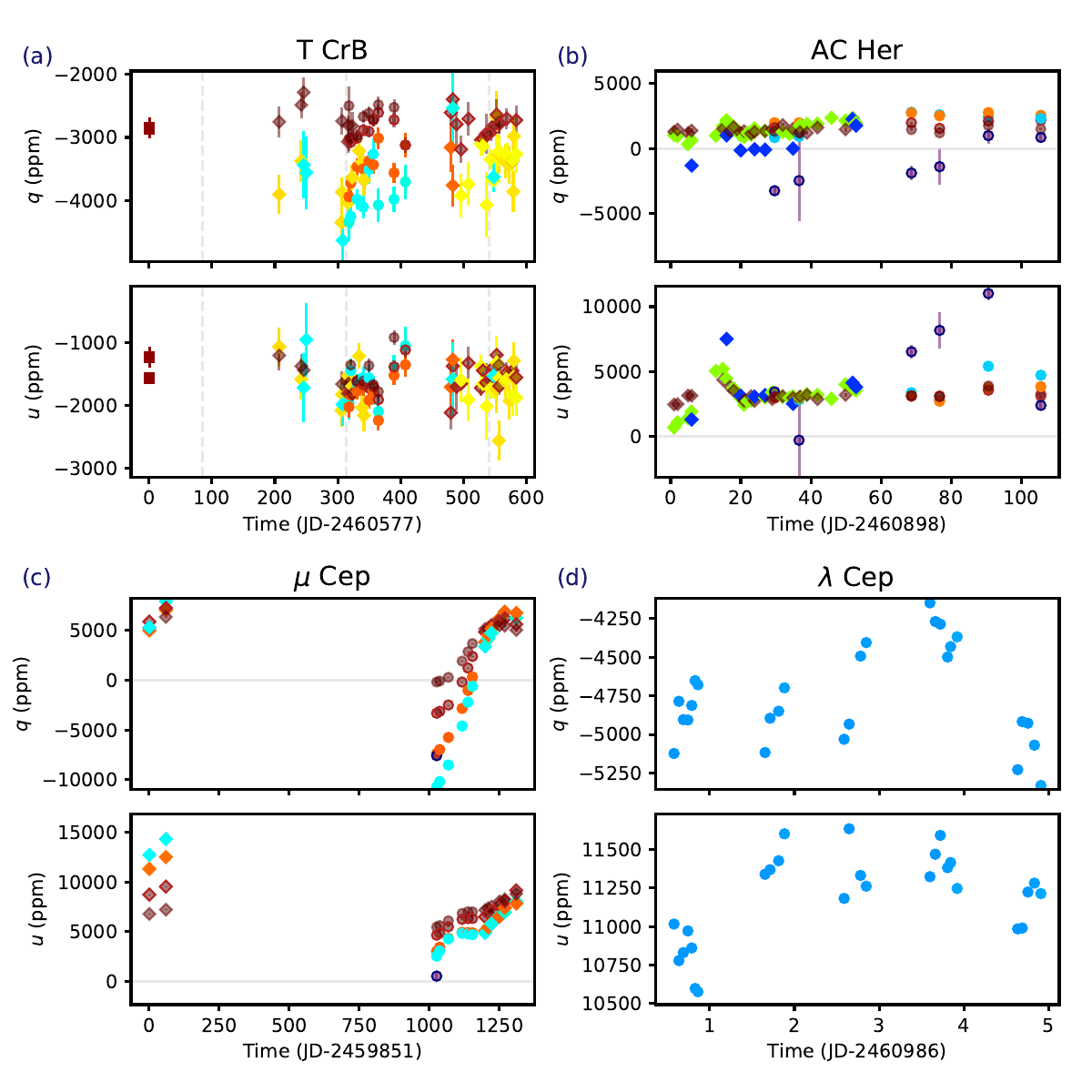}
\caption{Four variable stars observed with PICSARR-2 on MIRA's 36-inch telescope (circles), with PICSARR on the same telescope (squares), and PICSARR/-2 on smaller telescopes (diamonds). The colors represent $\lambda_{\rm eff}$ of the passband. Many of the observations of \mbox{T CrB} (a) and AC Her (b) have been made by one of us (JB)
with the 8-inch telescope at Pindari Observatory in Sydney, Australia. Some of these observations use non-SDSS bands, i.e. Chroma Bessel $B$ and $I$ bands, and a Luminance (L) filter from Astrodon, which is essentially $g^\prime+r^\prime$ -- these points show up as yellow in panel (a) and green in panel (b). A similar PlayerOne UV-IR Cut filter has been used on MIRA's C14, contributing the brighter yellow points in panel (b). To help distinguish $i^\prime$ from $z^\prime$, the former uses a red marker edge color, and the latter brown.}
\label{fig:other_var}
\end{figure*}

There are many other types of stars that make interesting objects for polarimetric study. The ability to make many observations over a long period of time is important in such work. In this regard the time available to small observatories and skilled amateurs is a significant advantage. Some examples are presented in Fig. \ref{fig:other_var}.

\subsubsection{T Corona Borealis}
\label{sec:T_CrB}

\mbox{T CrB}, the Blaze Star, is (currently) the faintest variable star we are observing at $m_V\sim10.5$. It is the brightest example of a recurrent nova with outbursts observed in 1866 and 1946 and another outburst expected soon \citep{Schneider2024}. Previous polarization observations \citep{Nikolov2022} showed interstellar polarization and no variability. In Fig. \ref{fig:other_var}(a) we see a convincing change in polarization for both visible and near-infrared passbands, with those in the near-infrared showing more stability. \mbox{T CrB} is a binary system ($P=227.57$ d) consisting of a red giant transferring material onto a white dwarf companion. The vertical grey dashed lines in Fig. \ref{fig:other_var}(a) correspond to the binary phase of the 1946 eruption ($E=2447918.66$ JD). We suspect that polarization variability may be caused by light reflected between the two binary components and if so this would allow us to determine the orientation of the binary orbit on the sky. This would then provide valuable information to help understand any structure seen in the outflowing gas when the outburst occurs.

\subsubsection{AC Herculis}
\label{sec:AC_Her}

AC Her is a member of the RV Tauri class of variables. These are solar type stars in their final stages of evolution. AC Her shows pulsations with a period of 75 days, and these are known to be accompanied with surprisingly large variations in polarization \citep{Henson1985, Nook1990}. The polarization amplitudes can be as large as 1 per cent (10000 ppm) at blue wavelengths and this is confirmed by our PICSARR observations in Fig. \ref{fig:other_var}(b). The large polarization variations of RV Tauri stars are something of a mystery, as most pulsating stars show only small polarization amplitudes (as described in Sec. \ref{sec:pulsating}).

\subsubsection{\texorpdfstring{$\mathrm{\mu}$}{mu} Cephei} 
\label{sec:mu_Cep}

First suspected variable by \citet{Hind1848}, and observed extensively since\footnote{Including a 55 year sequence by visual observer Joseph Plossman (the first part of the sequence can be found in \citealp{Plossman1904}).}, $\mathrm{\mu}$ Cep is an M2 supergiant classified as an SRc type, with prominent periods near 860 and 4400 d \citep{Brelstaff1997, Kiss2006}. Its polarimetric variability has also been known for a long time \citep{Grigorian1966}; it correlates, at least, to the shorter photometric period \citep{Polyakova2003}. The large amplitude, wavelength dependence, and slow polarimetric variability seen in Fig. \ref{fig:other_var}(c) is typical of this type of star. The most often invoked mechanisms relate to absorption and scattering from convective cells (hotspots) and/or circumstellar dust \citep{Schwarz1986}. Though a companion star within the extended dust shell has also been proposed to explain the same behavior in Betelgeuse \citep{Karovska1985, Karovska1986}, which is also an SRc type M2 supergiant.

\subsubsection{\texorpdfstring{$\mathrm{\lambda}$}{lam} Cephei} 
\label{sec_lam_Cep}

A program examining variability in early-type supergiants is nearing completion. It includes both stars that vary slowly like Deneb \citep{Cotton2024a} and stars that vary quickly like $\mathrm{\lambda}$ Cep \citep{Hayes1978}. The favored mechanism in these instances is a clumpy stellar wind that acts as an asymmetric scattering medium propagating outward from the star. Data from a five-night run is presented in Fig. \ref{fig:other_var}(d) that shows the star’s long-established night-to-night polarization changes, as well as faster variability. In addition to similar stochastic behavior, $\mathrm{\zeta}$ Pup, which has a similar spectral type, demonstrates a persistent polarimetric periodicity \citep{Bailey2024}. This speaks to the nature of the wind generation, and so it is a matter of interest as to whether similar behavior is present in $\mathrm{\lambda}$ Cep and other related stars.

\subsection{Imaging Polarimetry of NGC 7027}
\label{sec:NGC_7027}

\begin{figure}
\begin{center}
\includegraphics[width=0.98\columnwidth]{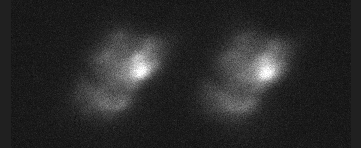}
\end{center}
\includegraphics[width=\columnwidth]{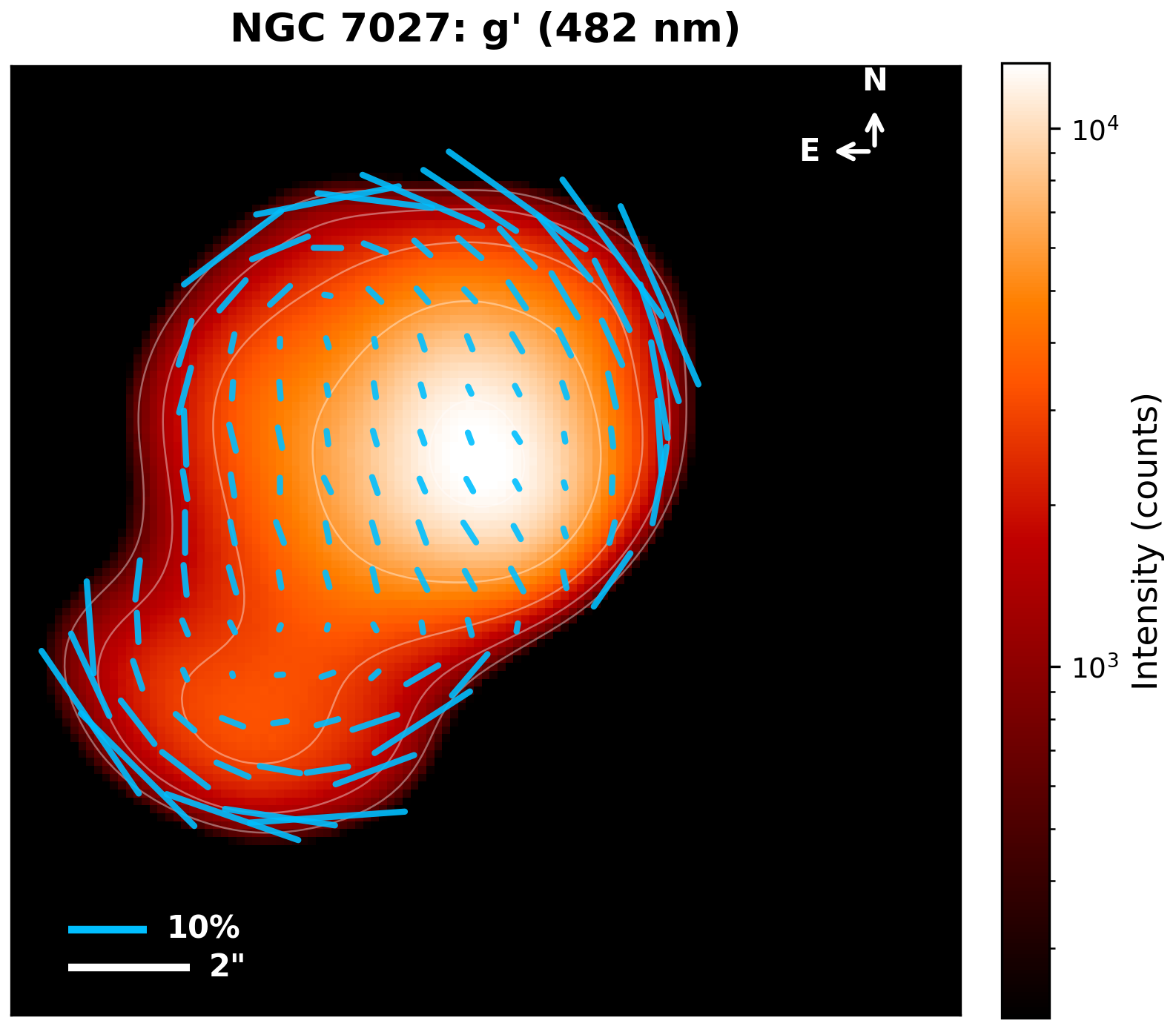}
\caption{Top: A single 500 ms frame dual-beam exposure of NGC 7027 in the $g^\prime$ filter; the detector was binned 2$\times$2. This is similar to what is seen during the observation using \textsc{SharpCap}. Bottom: A processed 20 minute $g^\prime$ exposure of NGC 7027 with PICSARR-2 on the 36-inch telescope, made up of the 500 ms frame exposures. The reduction scheme uses a 2D cross-correlation and whole-pixel shifts to stitch the frames together and obtain the polarization data. The color scale shows intensity in arbitrary units on a log scale; the blue vectors represent the polarization magnitude and orientation for a 3$\times$3 binned region. An intensity cut-off equal to 5 percent of the peak signal has been applied to the polarization map.}
\label{fig:NGC_7027}
\end{figure}

Post-AGB stars like AC Her (Sec. \ref{sec:AC_Her}) go on to form planetary nebulae. As an example of the imaging capability of PICSARR-2, we present an observation of NGC 7027 in the $g^\prime$ filter in Fig. \ref{fig:NGC_7027}. NGC 7027 is 10th magnitude, 600-year-old planetary nebula (PN) at a distance of $\sim$1 kpc, spanning $\sim$16-18 arcsec \citep{MoragaBaez2023}. After correction for interstellar polarization ($p=$ 0.822 percent, $\theta=$ 66.0 degrees) the intensity-weighted polarization (equivalent to a synthetic aperture measurement) is $p=$ 1.48 percent, $\theta=$ 154.3 degrees. The nebula displays a strong, centro-symmetric polarization pattern characteristic of dust scattering in Fig. \ref{fig:NGC_7027}, as it does in the 10.6 and 12.6 $\mu$m bands \citep{Jurgenson2003}, but with azimuthal (or tangential) rather than radial polarization. The polarization magnitude increases from $p=$0.63 $\pm$ 0.45 percent in the bright core to $p=$ 49 $\pm$ 25 percent in the faint outer regions (where the errors are derived from the standard deviation of pixels binned by intensity in these regions). These values agree well with the spectropolarimetric measurements of \citet{Walsh1994}, who using the \mbox{[O \textsc{iii}]} line (500.7 nm) found $p=$ 0.2 percent at the bright core and $p=$ 40 percent in the 20 arcsec halo.

We have acquired additional data at other wavelengths, which will be presented in a future work. Imaging polarimetry of nebulae in the optical is not as common as in the infrared. Being able to map the polarization wavelength dependence in these bands will enable a deeper exploration of the dust morphology.

\section{Discussion}
\label{sec:discussion}

Recent technological advances -- particularly the development of CMOS cameras for astronomy (refs. in \citealp{Bailey2023}) -- have allowed for new levels of efficiency and precision to be achieved from polarimeters built on the foundations of much older innovations in optics. Broadband stellar polarimetry has many applications, with some of the most exciting relating to the study of variable stars. The polarimeter described here is inexpensive, yet capable of accessing all of these polarigenic mechanisms with only a small or moderate sized telescope. This makes it ideal for use by the individual investigator or small teams.

Stellar polarization at the ppm level has been little explored. The most extensive survey is that of \citet{Piirola2020} of 361 stars, of which all but two are 4th magnitude or fainter. Around 100 stars mostly brighter than 3rd magnitude were measured in two other surveys \citep{Bailey2010, Cotton2016}. This leaves the bulk of naked eye stars without a high precision observation reported in the literature.

As shown in Sec. \ref{sec:sci_eg}, a PICSARR-like instrument is capable of efficiently investigating both low level polarimetric variability in bright stars, and higher levels in fainter stars. In the latter case \citet{Clarke2010}, in surveying the literature, highlights many cases where an interesting result has not been followed up. Furthermore, he gives a plethora of examples of variability studied only decades ago, now ripe for more detailed analysis with modern equipment. For smaller polarization variability examples, \citeauthor{Piirola2020}'s catalogue includes a list of 18 stars with repeat observations they consider variable (their Table 11); these remain uninvestigated. No doubt there are many more similar stars to be found. Indeed, there are many such lists in older catalogues.

Polarimetry also works well in tandem with other techniques, providing complimentary information. Recently, \citet{Bailey2024} combined small telescope ground-based polarimetry and spectroscopy with freely available photometric data from the \textit{Transiting Exoplanet Survey Satellite} (\textit{TESS}, \citealp{Ricker2015}) to provide new insights into 2nd magnitude $\zeta$ Pup. \citet{Cotton2024a} used the same combination to better understand Deneb's variability (the investigation is ongoing, \citealp{Osterhoudt2026}). There are considerable advantages to members of AAVSO, and other similar organisations, working together on these types of projects.

\section{Conclusions}
\label{sec:conclusions}

PICSARR-2 is a very versatile, yet inexpensive, linear polarimeter capable of synthetic aperture polarimetry and imaging to high precision and efficiency, over a broad magnitude range, on small to moderate sized telescopes. It is especially well suited to studies of stellar variability.

We have built a modified copy of the PICSARR-2 instrument and thoroughly tested it on MIRA’s 36-inch and C14 telescopes. Of the two cameras used, the ZWO ASI 462MM provides better performance on fainter stars, but the ZWO ASI 662MM is a better choice for bright star work. The instrument’s polarization precision is about a factor of 8 improved over the best reported for the original PICSARR. On the 36-inch telescope the limiting precision, $e_p$, is 1.3 ppm on bright stars -- making it one of the most precise stellar polarimeters in the World -- while the 5.8 ppm achieved on the C14 is as much as could be utilized for stellar astronomy on a small telescope. The position angle precision is also improved, being better than 0.1 degrees on the larger telescope.

On high polarization standard stars -- $\phi$ Cas, HD 23512, $o$ Sco, HD 154445, $\eta$ Aql -- the standard deviation observed in $p$ -- across the two telescopes and over about 300 nights -- is less than reported by \citet{Cotton2024b}. Whereas the variation in $\theta$ for the same stars matches well.

The main observing program for PICSARR-2 is focused in B-type pulsating stars. The instrument has proven suitable for this challenging work. A relatively large polarization signal attributable to non-radial pulsations is seen in $\mathrm{\epsilon}$ Per, but no significant polarization was detected in observations of $\mathrm{\zeta}$ Peg. The main polarization signal from $\mathrm{\zeta}$ Cas results from its weak magnetic field. Of the examples of joint programs with other PICSARR instruments on small telescopes, the highlight is polarization variability detected from \mbox{T CrB} for the first time. 

An example of imaging polarimetry with \mbox{PICSARR-2} of NGC 7027 in the $g^\prime$ band shows a strong centro-symmetric azimuthal polarization, increasing in strength from the bright center to the fainter outer regions.

\pagebreak

\acknowledgments

We thank the anonymous reviewer for their valuable feedback. Max Anthony Velasquez, Jean Perkins, Ievgeniia Boiko, Gerhard Gross, Himanshu Jain, August Davis, and Corinna Olson assisted with some of the PICSARR-2 observations used in the paper. Ain De Horta contributed some comparison observations made at Western Sydney University’s Penrith Observatory. Discussions with Derek Buzasi and Conny Aerts helped formulate the target list for the pulsating star program. A study by Richard Hubbard helped determine the best synthetic aperture radii to use. We would like to thank Robin Glover of SharpCap for his responsiveness in making changes to the program to suite PICSARR, and Matt Jennings of Thorlabs for his invaluable assistance organizing characterization of the AHWP05M-580 modulator. The adaption of PICSARR-2 for MIRA's 36-inch telescope is supported by NSF grant AST 2320626; observations of $\mathrm{\beta}$ Cep and SPB stars with the instrument are supported by NSF grant AST 2407635. We also thank the Friends of MIRA for their support.

\bibliography{bib}{}
\bibliographystyle{aasjournal}

\end{document}